\documentclass[a4paper,fleqn,usenatbib]{mnras}
\usepackage{newtxtext,newtxmath}
\usepackage[normalem]{ulem}

\usepackage[T1]{fontenc}
\usepackage{ae,aecompl}

\usepackage{graphicx}	
\usepackage{amsmath}	
\usepackage{amssymb}	

\title[Dynamical formation of BH LMXBs in the field]{Dynamical formation of Black Hole Low-Mass X-Ray Binaries in the field -- an alternative to common envelope}

\author[J. Klencki et al.]{
Jakub Klencki,$^{1}$\thanks{E-mail: jklencki@astrouw.edu.pl}
Grzegorz Wiktorowicz,$^{1}$
Wojciech G{\l}adysz$^{1}$
and Krzysztof Belczynski$^{1}$
\\
$^{1}$Astronomical Observatory, University of Warsaw, Al. Ujazdowskie 4, 00-478 Warsaw, Poland
}

\date{Accepted XXX. Received YYY; in original form ZZZ}

\pubyear{2016}

\begin{document}
\label{firstpage}
\pagerange{\pageref{firstpage}--\pageref{lastpage}}
\maketitle

\begin{abstract}
Very wide binaries ($> 500$ AU) are subject to numerous encounters with flying-by stars in the Galactic 
field and can be perturbated into highly eccentric orbits ($e \sim 0.99$). For such systems tidal interactions
at close pericenter passages can lead to orbit circularization and possibly mass transfer, consequently 
producing X-Ray binaries without the need for common envelope. We test this scenario for the case of 
Black Hole Low-Mass X-Ray Binaries (BH LMXBs) by performing a population synthesis from primordial binaries
with numerical treatment of random stellar encounters. We test various models for the threshold pericenter 
distance under which tidal forces cause circularization. 
We estimate that fly-by interactions can produce a current population of $\sim60$--$220$ BH LMXBs in the Galactic field.
The results are sensitive to the assumption on tidal circularization efficiency and zero to very small BH natal kicks of a few km/s
are required.
We show that the most likely donors are low-mass stars (< 1 $\rm M_{\odot}$; at the onset
of mass transfer) as observed in the population of known sources ($\sim 20$). 
However, the low number of systems formed along this route is in tension with 
most recent observational estimate of the number of dormant BH LMXBs in the Galaxy 
$10^4$--$10^8$ \citep{Tetarenko2016}. If indeed the numbers are so high, 
alternative formation channels of BHs with low-mass donors need to be 
identified. 

\end{abstract}




\section{Introduction}
There are currently 19 Galactic X-ray binary systems in which black holes (BHs) have been confirmed dynamically, 
with a few more extragalactic sources observed \citep{Remillard2006, Casares2014, Wiktorowicz2014, Tetarenko2016B}. 
They are all close binaries (with the majority of orbital periods $P_{\rm orb} < 1$ day), for which a mass transfer occurs and an accretion disc is formed
around the black hole and is responsible for the X-ray activity (\citealt{Shakura1973}; \citealt{Lasota2015}). 16 of these systems possess companion stars of 
relatively low mass ($<2 \,\rm M_{\odot}$), with the distribution peaking at around $0.6\, M_{\odot}$ and spectral types ranging from A2V to M1V
-- they are classified as low-mass X-ray binaries (LMXBs). All known BH LMXBs are transient X-ray sources, 
exhibiting occasional outbursts with their brightness increasing by 3-5 orders of magnitude, which are attributed 
to disc instability \citep[e.g.][]{Lasota2001}. 

The BH LMXBs are most commonly thought to originate from primordial binaries of a black hole progenitor \citep[$M_{\rm a} > 20-25 \, \rm M_{\odot}$, e.g.][]{Fryer2012}
and a much less massive main sequence (MS) secondary ($M_{\rm b} < 2 \, \rm M_{\odot}$) although evolution starting with intermediate mass secondaries was also
proposed \citep{Justham2006, Chen2006}. As the primary expands during its evolution it is expected to fill its Roche lobe and launch a
dynamically unstable mass transfer, initiating the common envelope (CE) phase \citep{Paczynski1976}. As a result of this short-lived evolutionary stage the separation 
between the components decreases significantly, which ultimately leads to a later mass 
transfer from low-mass star to a black hole and prolonged X-ray activity.
The formation of a black hole from a Zero-Age Main-Sequence (ZAMS) star takes less than 10 Myr. In such a time the low-mass secondary is still at the beginning
of the MS \citep[or even still during pre-MS contraction;][]{Ivanova2006} and is expected to remain a MS star for the next $2.5$--$10$ Gyr.

This now considered standard scenario was first suggested nearly 30 years ago to explain the origin of the A0620-00 black hole X-ray binary \citep{deKool1987}. Since then, however, multiple 
studies have pointed out flaws of this conception in the particular case of BH LMXBs, suggesting that it is difficult for a binary of a massive giant and a much less massive 
companion to eject the CE at the expense of its orbital energy \citep[e.g.][]{Podsiadlowski2003}. 
Thus, unless abnormally high values for the CE ejection efficiency are adopted \citep{Yungelson2008}, the system is most likely to merge. In an attempt to justify such 
increased efficiency several modified CE models were proposed (see \citealt{Li2015} and references therein). Population synthesis studies have recently shown, however, that 
CE models in their current state fail to reproduce the distribution of donor mass of the observed BH LMXBs (\citealt{Wiktorowicz2014}; \citealt{Wang2016}), implying the need for 
either a more significant adjustment of the CE modeling or an alternative formation scenario to be considered.

Recently \citealt{Michaely2016} (hereafter MP2016) suggested that LMXBs 
(including those with black holes) could originate from very wide binaries ($a > 1000\,\rm AU$) that undergo a series
of subsequent fly-by interactions with stars of the Galactic field. The authors 
predict that some of these wide systems should be excited into orbits of very high eccentricity, eventually circularized by tidal effects at pericenter. This 
essentially transforms a wide binary into a compact one, playing a similar role to the CE phase in the standard scenario. 

Dynamical formation of LMXBs is definitely not a new concept when it comes to dense stellar environments, such as globular clusters, where gravitational interactions and tidal captures 
take place most often (\citealt{Clark1975}; \citealt{Fabian1975}). In fact, it appears that formation channel of X-ray binaries through dynamical processes is very efficient, as
$\sim 10\%$ of all LMXBs is being found in globular clusters, which contain only about $\sim 0.1 \%$ of all stars in the Galaxy \citep{Irwin2005} (note that these are all X-ray binaries with a 
neutron star as the accretor as all known BH LMXBs reside outside globular clusters). 
Similar processes 
should be effective also in the most dense, central regions of galaxies. This was confirmed by \citet{Voss2007} who carried out Monte Carlo simulations of a binary population
in the bulge of M31, applying the \textsc{fewbody} code \citep{Fregeau2004} to numerically simulate random dynamical interactions with passing by stars. 

A similar study, however, has never been done for the Galactic disc, where all of the known BH LMXBs reside \citep{Li2015}. Although the disc is typically considered to 
not be dense enough for stellar encounters to play an important role, dynamic interactions do become significant for wide binaries with naturally larger cross-sections (e.g. \citealt{Bahcall1985}).
It is even predicted that very wide binaries should be the primary source of stellar collisions in the Milky Way \citep{Kaib2014}, thanks to similar excitations to highly eccentric 
orbits as considered by MP2016 in their scenario for LMXBs formation.

Here we perform a population synthesis study on the dynamical formation of BH LMXBs from primordial binaries in the Galactic disc. We focus only on binaries with a black hole as their primary
and a MS star as their secondary, as the origins of these systems remain most mysterious \citep[MS companions are also most frequent among observed systems, as $\sim75\%$ of donors are 
low-mass dwarf stars,][]{CorralSantana2016}.

The paper is organized as follows: in section 2 we describe the MP2016 scenario, as well as introduce our 
modeling of dynamic interactions and tidal circularization. In section 3 we describe our population synthesis approach and utilized computer codes. 
In section 4 we present and discuss our results. We conclude in section 5. 

\section{Dynamical interactions as source of X-ray binaries}

\subsection{LMXB formation scenario of MP2016}
The scenario for the dynamical formation of a LMXB is explained in details in MP2016, with an analytical description 
for the probability of the MS companion being tidally captured by the BH in a BH-MS system due to an interaction with a field star given therein. 
Here we briefly present the main points as well as provide a model for integrating dynamical interactions into binary population synthesis, 
essentially substituting the approximate analytical approach of MP2016 with the numerical treatment of stellar encounters.

Let us consider a wide binary ($a > 1000 \rm AU$) of a stellar BH and a companion MS star. Its large cross-section makes it subject to 
random short duration interactions with flying-by field stars. Such encounters affect the system's angular momentum and energy, 
thus changing its orbital parameters. MP2016 predict that it is possible for a binary to be perturbated into a very highly eccentric 
orbit ($e \approx 1$), for which the distance between components at pericenter ($d_{\rm per} = a(1-e)$) can be very small. 
They define a threshold distance $d_{\rm tidal}$ between components, at which tidal effects become strong enough to cause
circularization of the orbit -- a wide, eccentric BH-MS system fulfilling $d_{\rm per} < d_{\rm tidal}$ is thus transformed by tidal 
forces into a close, circular BH-MS binary. Such a system is subject to processes which may further decrease its separation (e.g. 
magnetic braking, gravitational waves radiation) leading to mass transfer and consequent formation of a LMXB.

\begin{figure}
	\includegraphics[width=\columnwidth]{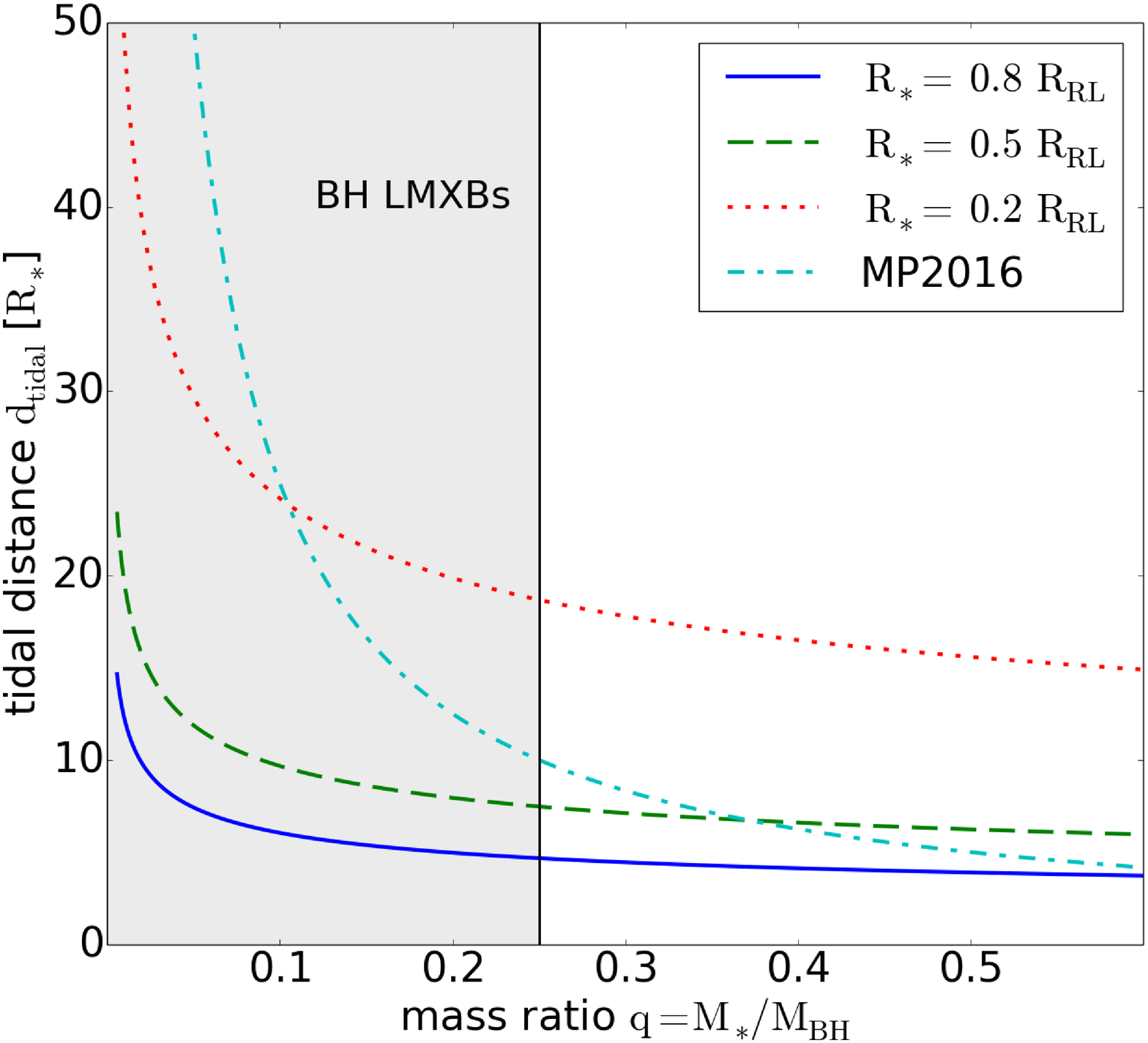}
    \caption{Comparison of 4 different models for the tidal distance $d_{\rm tidal}$ between components in a BH-MS binary, at which we assume tidal circularization becomes 
    significant. $\rm d_{\rm tidal}$ is expressed in the unit of radius of the MS component $\rm R_*$. 
    The first 3 models are based on the minimal required Roche Lobe filling factor $\alpha$, i.e.: $R_{*} = \alpha R_{\rm RL}$, with the values $\alpha = 0.8$ 
    (most realistic), $\alpha = 0.5$ (optimistic) and $\alpha = 0.2$ (very optimistic). The last model is following relation~\ref{eq:mp2016} proposed by MP2016.
    }
    \label{fig:d_tidal_models}
\end{figure}

Estimating the value of $d_{\rm tidal}$ for a given binary is a complicated problem since our understanding of tides and energy dissipation mechanisms remains very uncertain. For their 
model MP2016 adapted results obtained by \citet{Kaib2014} who in the regime of dynamic tides investigated timescales of circularization due to energy dissipation at pericenter passages 
for equal mass MS-MS wide eccentric binary systems. They conclude that circularization becomes significant (i.e. energy dissipated during single pericenter passage is 
comparable with the entire orbital energy)
for $d_{\rm tidal} \approx 5 R_{\rm mean}$, where $R_{\rm mean}$ 
is a mean radius of the binary components -- in the case of a BH-MS system $R_{\rm mean} = 0.5 \, R_{*}$, where $R_{*}$ is the radius of the MS secondary. 
This result is in agreement with prescription developed by \citet{PortegiesZwart1996} stating that circularization occurs immediately if the stellar radius of one
component is larger than $0.2$ of the distance between binary components at periastron.
In order to adopt the result of \citet{Kaib2014} to a binary of unequal masses MP2016 apply an additional factor of mass ratio $1/ q = M_{\rm BH} / M_{*}$:
\begin{equation}
 d_{\rm tidal\_MP} = 2.5 \; \; R_{*} \; \frac{M_{\rm BH}}{M_{*}} = 2.5 \; \; R_{*} \; \frac{1}{q},
\label{eq:mp2016}
\end{equation}
\noindent where $M_{\rm BH}$ and $M_{*}$ are masses of BH and MS respectively. Adding the term $1/q$ alone is a very simplified approach to stellar mass scaling (as it does not 
take into account changes of stellar radius or inner structure), which can be often justified in problems we have poor understanding of, such as tidal interactions. 
However, according to relation (1) the ratio $d_{\rm tidal} / R_{*}$ grows very big for small values of $q$ expected in the case of potential BH LMXBs progenitors, 
as a typical mass of a stellar Galactic BH is around $7$--$8 \rm M_{\odot}$ and the MS companion is a low-mass star, often $< 1 \rm M_{\odot}$ \citep{Casares2014}.
We find this behavior rather controversial 
and decide to additionally consider models which are less dependent on the components mass ratio, 
for which the threshold distance between binary components $d_{\rm tidal}$ is estimated based on the ratio $\alpha$ of
MS radius and the size of its Roche Lobe (Roche Lobe filling factor), i.e.:
\begin{equation}
 R_{*} = \alpha \; R_{\rm RL} = \alpha \; \; q_{\rm RL} \; \; d_{\rm tidal}
\end{equation}
\noindent where $q_{\rm RL}$ can be approximated with the formula of \citet{Eggleton1983}:
\begin{equation}
 q_{\rm RL} = \frac{0.49 q^{2/3}}{0.6 q^{2/3} + \ln(1 + q^{1/3})}
\end{equation}
\noindent This yields the condition for $d_{\rm tidal}$ to be:
\begin{equation}
d_{\rm tidal} = \alpha^{-1} \; \; R_{*} \; \; q_{\rm RL}^{-1}
\end{equation}
In our population synthesis we test one model following relation~\ref{eq:mp2016} proposed by MP2016 as well as three models corresponding to values $\alpha = 0.2$ 
(very optimistic), $\alpha = 0.5$ (optimistic) and $\alpha = 0.8$ (most realistic) in relation (4).
Fig.~\ref{fig:d_tidal_models} shows a comparison of $d_{\rm tidal}$ vs $q$ relation in each of these cases. 
The expected range of $q$ for LMXBs is about $q < 0.25$.

For a mass ratio $q = 0.2$ the size of a MS secondary Roche Lobe is approximately equal to $0.25$ the distance between binary components (i.e. $R_{\rm RL} \approx 0.25 \; d_{\rm tidal}$).
This means that for $q \gtrsim 0.2$ the model with $\alpha = 0.8$ (i.e. $ R_{*} = 0.8 \; R_{\rm RL}$ required for circularization) is roughly corresponding to the 
prescription for immediate circularization
$R_{*} = 0.2 \; d_{\rm tidal}$ given by \citet{PortegiesZwart1996} and supported by \citet{Kaib2014}. For lower mass ratios, however, especially in our optimistic models,
the threshold pericenter distance required for tidal circularization corresponds to $R_{*} < 0.2 \; d_{\rm tidal}$. In such case the timescale of 
circularization could be significantly longer, possibly even a few Gyr. 

\begin{figure*}
	\includegraphics[width=450px]{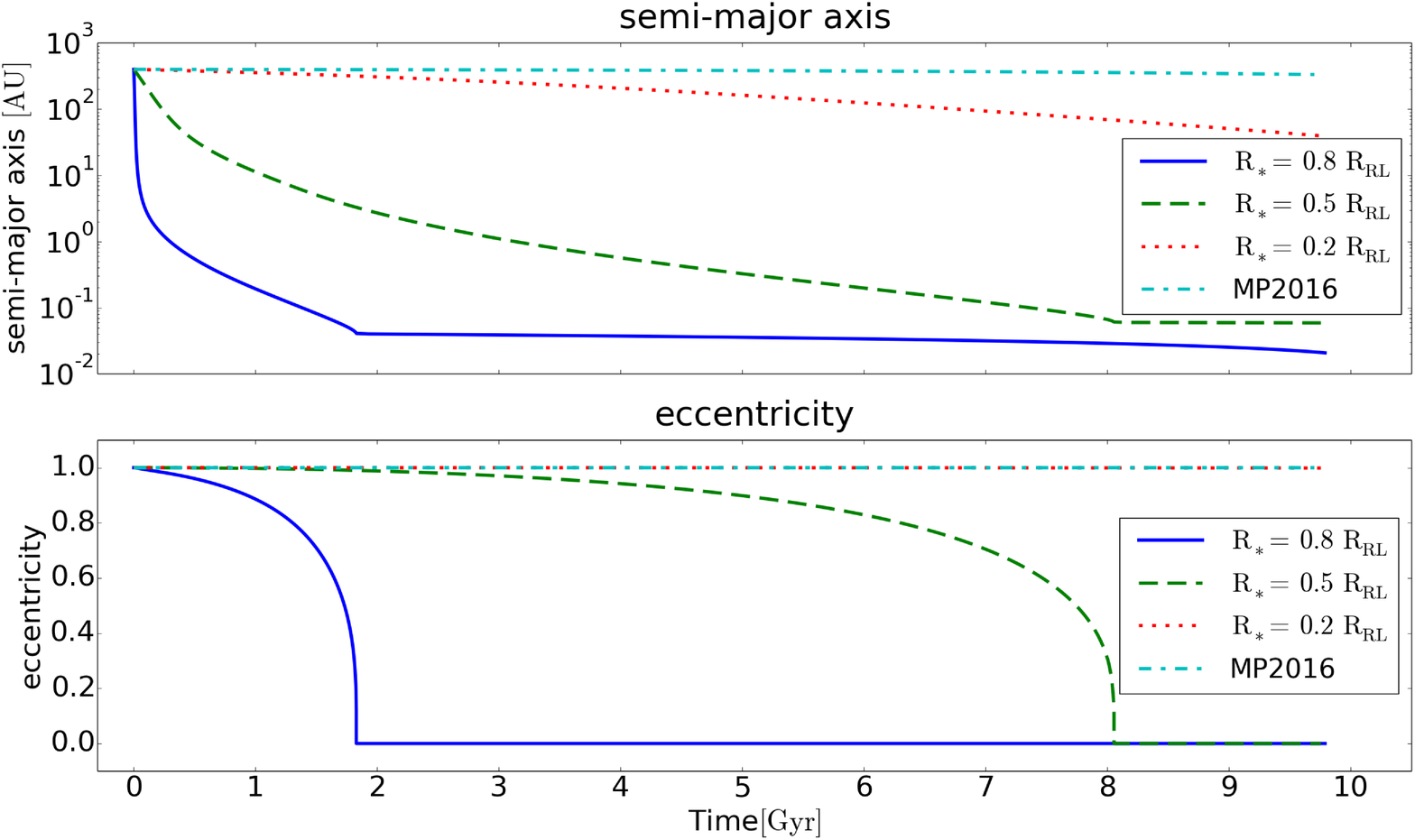}
    \caption{Evolution of orbital parameters as calculated by \textsc{StarTrack} of a wide and eccentric BH-MS
    binary  (initial parameters: $\rm a_0 \approx 400 \; AU$ and $\rm e_0 \gtrsim 0.999$) fulfilling threshold condition 
    for tidal circularization in 4 of our tested models. The binary components masses were $\rm M_{BH} = 8.3 \rm \; M_{\odot}$ and $\rm M_{MS} = 0.87 \rm \; 
    M_{\odot}$, the MS star radius was  $R_* = 0.78 \rm \; R_{\odot}$ at the beginning, evolving up to $0.79 \rm \; R_{\odot}$ at 2 Gyr and 
    $0.85 \rm \; R_{\odot}$ at 8 Gyr. The value of eccentricity was set such for each model that the 
    exact threshold condition for circularization was satisfied ($e$ between $0.99978$ and $0.999946$ across the models).}
    \label{fig:tidal_circularization}
\end{figure*}

To illustrate this we simulated the evolution of orbital parameters ($a$,$e$)
of an exemplary wide and eccentric BH-MS binary from our population synthesis modeling using the \textsc{StarTrack} binary evolution code \citep[][detailed 
description in section~\ref{sec:startrack}]{Belczynski2002,Belczynski2008} -- see figure~\ref{fig:tidal_circularization}. \textsc{StarTrack} calculates 
tidal evolution in the equilibrium-tide, weak-friction approximation of \citet{Zahn1989}, using \citet{PortegiesZwart1996} prescription for immediate 
tidal circularization. The system of interest consists of a $8.3 \rm \; M_{\odot}$ BH and a $0.87 \rm \; M_{\odot}$ secondary MS of radius $R_* = 0.78 \rm \; R_{\odot}$. 
The starting semi-major axis was $a = 390 \rm \; AU$, whereas the value of eccentricity was set such that for each model the exact threshold condition for circularization 
was satisfied ($e$ between $0.99978$ and $0.999946$ across the models). For such a low mass ratio $q = 0.87 / 8.3 \approx 0.1$ the condition $R_{*} = 0.2 \; d_{\rm tidal}$ 
is never fulfilled, thus circularization does not occur instantaneously as calculated in \textsc{StarTrack}. 
In fact, as figure~\ref{fig:tidal_circularization} shows, only in models with $\alpha = 0.5$ and $\alpha = 0.8$ the orbital size decreased quickly enough and 
full circularization was complete in under 10 Gyr. Even though the model with $\alpha = 0.2$ and the one adopted from MP2016 are thus proven to be 
extremely optimistic in the tidal regime of \citet{Zahn1989}, we decide to test them regardless in order to investigate how the efficiency of dynamical 
formation channel for BH LMXB could be affected by exceptionally strong tidal forces. 

\subsection{Dynamical interactions in the field}
\label{sec:dynamic_theory}
In this study we investigate a simplified scenario in which wide binaries encounter only single stars, i.e. we assume the
binary fraction of passing by stars equal to zero and only consider binary-single 
interactions. Let us consider a binary of masses $M_1$, $M_2$ and a semi-major axis $a$, located in the Milky Way at position $\vec{r}$
and moving with velocity $\vec{v}$. With its position in the Galaxy we can associate a value of local stellar number density $n_{*}$ and 
a local distribution of stellar velocities $f(\vec{v_{*}})$. The expected number of encounters in a time frame $\Delta t$
with stars passing by within distance $b_{\rm max}$ from such binary can be expressed as:
\begin{equation}
 N(\Delta t) = n_{*} \; \Delta t \; \pi b_{\rm max}^2 \; \int |\vec{v_*} - \vec{v}| \; f(\vec{v_*}) \; d\vec{v_*}
 \label{eq:NOI}
\end{equation}

We define mean free time $t_{MF}$ as a mean time between stellar encounters for such a binary, which is equal to:

\begin{equation}
 t_{\rm MF} = \frac{\Delta t}{N(\Delta t)}
\end{equation}
The value of $b_{\rm max}$ was chosen as $b_{\rm max} = 5a$ because for larger impact parameters 
the eccentricity changes induced by a fly-by are negligible \citep[$\delta e / e \ll 1$;][]{Fregeau2004,Heggie1996}. It is worth noticing here that the rate of encounters ($t_{\rm MF}^{-1}$) 
is greater not only for wider binaries ($t_{\rm MF}^{-1} \propto a^{2}$) but it also increases for systems moving with velocity significantly different from the local 
mean velocity of passing-by stars. 

Mean free time is a useful quantity for Monte Carlo simulations of a binary evolution. Upon each time step $\delta t \ll t_{\rm MF}$ we randomize whether or not 
a flying-by field star was encountered during this time, with the probability of such event equal to $\delta t / t_{\rm MF}$. If a star was encountered, we 
draw its relative velocity $\vec{v_{\rm rel}} = \vec{v_*} - \vec{v} $ from the local distribution of velocity $f(\vec{v_*})$, whereas its mass $M_{\rm single}$
is drawn from the Initial Mass Function (IMF). We use the Milky Way velocity distribution based on observational results of the Sloan Digital Sky Survey 
\citep[SDSS,][see section~\ref{sec:MW_dynamics} for more details]{Bond2010} and the IMF of \citet{Kroupa1993} with the power-law exponent modified to $-2.3$
for the stars of mass $\rm M> 1.0 \, M_{\odot}$, as recently adjusted by \citet{Sana2012}.

Another important quantity describing a binary-single interaction is impact parameter $b$. Due to effect of gravitational focusing, 
the cross-section for two masses $M_{\rm binary}$ and $M_{\rm single}$ passing with relative speed $v_{\rm rel}$ within an impact 
parameter $b$ from each other is given by the conservation of energy and momentum \citep[e.g.][]{Voss2007}:

\begin{equation}
\label{eq:cs}
 \sigma(b,v_{\rm rel}) = \pi b^2 \left( 1 + \frac{2G(M_{\rm binary} + M_{\rm single})}{b \;v_{\rm rel}^2}\right)
\end{equation}
Knowing $v_{\rm rel}$ and $M_{\rm single}$ we draw $b$ from the following distribution, obtained simply as a differential of the cross section~\ref{eq:cs}  
normalized to $b \in [0, b_{max}]$:
\begin{equation}
f(b,v_{\rm rel}) = \frac{1}{ \sigma(b_{\rm max},v_{\rm rel})} \frac{\partial \sigma(b,v_{\rm rel})}{\partial b}
\end{equation}

\section{Population synthesis modeling}

\subsection{General approach}

Our intention is to provide a realistic approach to the problem of  
dynamical interactions of wide binaries and single stars in the Galactic field.
We combine realistic initial conditions for stars and binaries, invoke
relevant binary processes (BH formation with its potential effects on binary
orbits) and apply a wide range of criteria on the key issue: tidal
circularization of hyper-eccentric binaries. 
Investigating the effect of three-body encounters on a system evolution is a computationally heavy task, 
since for every wide system one can expect on the order of thousands of dynamical interactions during its several Gyr 
long evolution, each of those needed to be simulated individually. To simplify the computations and focus on the 
most important factors for LMXBs formation in the MP2016 scenario we have split each binary evolution into two 
subsequent evolutionary stages: pre and post BH formation from the massive primary (hereafter denoted as 
the ''pre-BH'' and the ''post-BH'' stage). 

The pre-BH stage begins with a zero-age binary and ends once the primary has ended 
its evolution and formed a BH. Since it only lasts for up to 10 $\rm Myr$ (progenitors of stellar BHs evolve very quickly) 
we assume no dynamical interactions taking place during this stage. Even though for a very wide 
binary ($a \sim 10^4 \rm AU$) it is possible to obtain mean free time between interactions of the order of 
$\rm Myr$ it is unlikely that a single encounter will change orbital parameters in a significant way, thus affecting
the pre-BH evolution. 
This could happen if the encountered star was an exceptionally massive object of at least few solar masses (see Sec.~\ref{sec:ex_ev_route}
for discussion of this matter in the case of BH-MS systems). However, due to the combined effects
of stellar evolution and the IMF being shifted towards lower masses,
the majority ($> 90 \%$, see discussion in Sec.~\ref{sec:mass_func_flybys}) of field object are low-mass stars or white dwarfs ($M \lesssim \rm 2 M_{\odot}$).
On the other hand, internal binary evolution during this stage is very dynamic and thus the 
main focus of our simulations.

The post-BH stage begins with a newly-formed BH-MS binary and lasts until the present Galactic age.
The primary factor affecting a wide binary evolution during such a long period is the influence of dynamic interactions 
with fly-bys. This stage is to last several Gyr on average, the exact length depending on the moment of a binary's birth, 
thus multiple encounters with passing-by stars are expected and have to be accounted for.
Their frequency depends on the local conditions in the 
Galaxy (i.e. stellar number density, velocity distribution), thus the evolution of a binary position in 
the Milky Way potential has to be calculated as well. 

\subsection{Pre-BH stage -- binary evolution modeling}
\label{sec:startrack}

We employ the \textsc{StarTrack} binary population synthesis code to generate 
a population of BH-MS binaries. The code was created based on single stellar models adopted 
from \citet{Hurley2000} with special attention being given to massive stars and compact objects. 
A detailed description of \textsc{StarTrack} can be found in \citep{Belczynski2002} and 
\citep{Belczynski2008}. Factors especially important for the evolution leading to wide BH-MS binaries have recently been updated, 
including the addition of new prescription for wind mass loss as well as the implementation of
convection driven, neutrino enhanced supernova engines with rapid explosion development \citep{Dominik2012,Belczynski2012}. 
For BHs natal kicks during supernova explosion we adopt Maxwellian
distribution of kicks with $\sigma = 265 \rm \, km \, s^{-1}$ (as for neutron stars, see \citealt{Hobbs2005}),
with the value of the kick being lowered proportionally to amount of material falling back onto the BH \citep{Fryer2012,Belczynski2016}.
As a result the natal kicks for BHs are often negligible or very small . However, even in the case of no natal kick
each system evolving into BH-MS is given an additional velocity of $\sim 10-20\rm \; km \; s^{-1}$ due to Blaauw kick 
\citep{Blaauw1961}, which is associated solely with the mass loss of the primary during BH formation. 

For massive primaries we have chosen the mass range 19--150 $\rm M_{\odot}$, to ensure a BH formation, 
whereas for secondaries a corresponding 0.08$\rm \; M_{\odot}$--$\rm M_{\rm primary}$ range.
We adopt the initial parameters distributions from \citet{Sana2012}, implied from the spectroscopic measurements of massive O-type 
stars. These include: the IMF for the primary proposed by \citet{Kroupa1993} with the power-low exponent modified to $-2.3$
for the stars of mass $\rm M> 1.0 \, M_{\odot}$, 
binary mass ratio $q = \rm M_{\rm b} / \rm M_{\rm a}$ flat distribution
\citep[as also obtained by][]{Kobulnicky2006}, eccentricity distribution $f(e) \propto e ^{-0.42}$ in range $[0.0 \, , \, 1.0]$
and period distribution $f(P) \propto \rm log(P/day)^{-0.55}$ in range $[0.15 \, , \, 5.5]$. 
Note that spectroscopic measurements can only reliably detect binary systems of orbital periods up to $\rm log(P/day) \lesssim 3.5$. 
However, basing on recent interferometric observations of Galactic massive stars by \citet{Sana2014},
probing binaries of separations up to $\sim 200 \rm \; AU$,
we expand the period distribution up to $\rm log(P/day) \lesssim 5.5$
(similarly to \citet{deMink2015, Belczynski2016_nature} and other population synthesis works).

Such initial conditions result in widest binaries having separations of about $\sim 400 \rm \; AU$.
Meanwhile some companions to massive stars are visually detected on even much wider orbits, of the order of thousands AU,
thanks to lucky imaging \citep{Peter2012} and space-based observations \citep{Caballero2014,Aldoretta2015}.
However, these observations also reveal that many of such ultra-wide systems with massive stars are in fact triples.
In a recent work \citet{Moe2016} analyzed results of more than twenty surveys and
concluded that companions to massive stars with orbital periods of $\rm log(P/day) \gtrsim 5.5$ are essentially
tertiary components in hierarchical triples.
For such systems the dynamics during close pericenter passages 
of the wide component would be much more complex, making them unlikely progenitors of X-ray systems in the MP2016 scenario
-- see Sec. 3.1.4 of MP2016 for details. 
We thus consider $\rm log(P/day) = 5.5$ to be a reasonable upper cutoff of the orbital period of our starting binaries.

We have evolved $4.5 \times 10^7$ binaries starting from ZAMS, with stellar
formation history (SFH) assumed constant throughout the whole Milky Way age $t_{\rm disc} = 10 \; \rm Gyr$.
Given ${\rm SFR} = 3.5 \rm M_{\odot} yr^{-1}$ and binary rate $f = 0.5$ our sample corresponds to
about $94 \%$ of the binaries with $\rm M_{a} \geq 19 M_{\odot}$ in the Galactic disc population, 
so only a mild scaling was needed (i.e. the numbers 
of BH LMXB we eventually obtain accross all our models are on average $1.0 / 0.94 \approx 1.06$ times the numbers of actual 
BH LMXB systems in our simulations). 
We adopt Solar metallicity \citep[$\rm Z = 0.02$][]{Villante2014} to all stars in the Galactic disk that we evolve.

Additionally, in order to be able to do a straightforward comparison with the results of MP2016, we also 
evolved $1.215 \times 10^7$ ultra-wide binaries with initial periods of $\rm log(P/day)$ from range $[5.5 \, , \, 8.5]$ 
and $f(P) \propto \rm log(P/day)^{-0.55}$ (keeping \citet{Sana2012} distributions for other initial binary parameters).
Note that due to likely triple nature of systems this wide \citep{Moe2016} we exclude them from our main results
and only address in the discussion (\ref{sec:discussion}).

An exemplary BH-MS candidate for future dynamically formed BH LMXB in our simulations was born at 
ZAMS as a MS-MS system of $29 \rm \; M_{\odot}$ primary and $0.75 \rm \; M_{\odot}$ secondary, with 
orbital parameters $a = 92 \; \rm AU$ and $e = 0.51$. As a core helium burning giant the primary expands 
up to a radius $\sim 1730 \rm \; R_{\odot} \approx 8 \; AU$. However, since 
the separation between components is very high, the giant does not fill its Roche lobe and
no mass transfer occurs (in particular there is no CE phase). The primary loses the majority of its mass 
due to stellar winds, which in turn increases the semi-major axis. 
Throughout the pre-BH evolution the 
primary reduces its mass down to $\sim 8.8 \rm \; M_{\odot}$ and eventually becomes a
naked helium star. At this stage the semi-major axis has been increased up to $a = 290 \rm \; AU$, whereas 
the secondary mass and the orbit eccentricity remain unchanged. At around $6.8$ Myr since ZAMS the primary 
evolves into a BH through a core-collapse supernova. No direct natal kick has been given to BH since
all expelled material fell back onto it. Neutrino emission reduces its mass down to the final 
$\rm M_{BH} \sim 7.9 \rm \; M_{\odot}$, however, the associated Blaauw kick gives the binary only a negligible velocity 
of $0.03 \rm \; km \; s^{-1}$. The post-BH formation orbital parameters are $a = 306 \rm \; AU$ and $e = 0.47$.

It should be noted here that in our simulations we also evolve systems with much shorter separations $a < 10 \rm \; AU$ at ZAMS. 
During evolution as helium burning giant the massive primary in such binaries expands overflowing its Roche lobe and leading 
to the CE phase. In the vast majority of our systems this leads to a merger, as it is very difficult for a low-mass secondary to 
eject a massive envelope created by his companion \citep[e.g.][]{Wiktorowicz2014}. The few systems which have undergone and survived 
the CE phase were all close binaries with separations of the order of a few $\rm AU$. Due to their small cross-sections for stellar encounters 
they did not pay any role in the dynamical formation channel we investigated. 

\subsection{Post-BH stage -- stellar encounters modeling}

All successfully formed BH-MS binaries are made subject to dynamical interactions during the long
post-BH evolutionary stage. Similarly to the pre-BH stage and \textsc{StarTrack} code regime, 
we take Monte Carlo approach to model random stellar encounters. 

In this study we limit ourselves to BH-MS X-ray binaries only, which means that we require
the secondary star to remain at the main sequence for the whole 
duration of the post-BH stage and to not evolve onto the giant branch until the present day. This limitation
essentially favors the study towards less-massive companions (as they 
stay on the MS the longest), which are not only more numerous but, more importantly, are subject to larger 
perturbations from flying-by stars due to their lower orbital energy, thus being more likely to form 
an LMXB through the MP2016 scenario. It is worth pointing out here that the majority of donors in observed BH LMXBs 
are low-mass MS dwarfs \citep{Li2015}.

The big advantage of restricting ourselves to MS companions only is that low-mass MS stars evolve very slowly, without 
changing their mass or radius in a significant manner. This allows us to assume no stellar 
evolution throughout the post-BH stage. Same assumption can 
be made when it comes to interactions between binary components -- such as tidal forces \citep[eg.][]{Zahn1989}, 
mass transfer, gravitational waves emission \citep{Peters1964} or magnetic braking \citep[eg.][]{Rappaport1983}. 
In the case of binaries comprised of a stellar BH and a low-mass MS companion 
those effects only become important for systems with separations of a few solar radii. Meanwhile, 
encounters with field stars become relevant for binaries with semi-major axes of at least 10--100 AU (due to their cross section), 
which is a few orders of magnitude more. Of course once a condition for tidal circularization is fulfilled and 
a wide and eccentric BH-MS binary becomes a compact and circular system with a separation of a few solar radii then the above-mentioned 
effects become significant again. However, the evolutionary fate of systems post tidal circularization is subject to large uncertainties 
(see the discussion in Sec.~\ref{sec:tidal_circularization}) and we do not model it in this study.

\subsubsection{Binary dynamics evolution}
\label{sec:MW_dynamics}

Each system is first assigned with a certain position inside the disc, as well as 
given its disc velocity drawn from the local velocity distribution, which is added to the velocity acquired during its
pre-BH stage evolution ({\textit{i.e.} due to Blaauw kick plus potentially non-zero natal kick). For the 
Milky Way disc model we have adopted distributions fitted to the observational results of SDSS survey: the stellar number density distribution 
from \citet{Juric2008} and stellar velocity distributions from the kinematics study of \citet{Bond2010}. 

Following \citet{Gilmore1983} we model the disc as a sum of two exponential components 
(the ''thin'' and the ''thick'' disc) allowing for different scale lengths ($L_{\rm thin}$ and $L_{\rm thick}$) and 
heights ($H_{\rm thin}$ and $H_{\rm thick}$) for each component.

\begin{equation}
 \rho_{D}(r,z) = \rho_{D}(r,z;L_{\rm thin},H_{\rm thin}) + f\rho_{D}(r,z;L_{\rm thick},H_{\rm thick})
\end{equation}
where
\begin{equation}
 \rho_D(r,z;L,H) = \rho_D(r_{\odot},0)e^{\frac{r_{\odot}}{L}} {\rm exp} \left( -\frac{r}{L} - \frac{z + z_{\odot}}{H} \right)
\end{equation}
We adopt bias corrected scales of \citet{Juric2008}: 
$L_{\rm thin} = 2600 \rm \, pc$, $H_{\rm thin} = 300 \rm \, pc$ for the thin disc and $L_{\rm thick} = 3600 \rm \, pc$, 
$H_{\rm thick} = 900 \rm \, pc$ for the thick one. The less massive thick disc is normalized to $f = 0.12$ with
respect to thin disc, normalized to $1.0$. We constrain the size of the disc with $|z_{\rm max}| = 4000 \rm \, pc$,
$r_{\rm min} = 3000 \rm \, pc$ and $r_{\rm max} = 15000 \rm \, pc$, thus staying outside of the influence of central bulge, 
for which the results of SDSS are uncertain due to high blending. For the value of stellar number density in the Sun neighborhood
we adopt $\rho_{D}(r_{\odot},0) = 0.005 \, \rm ly^{-3}$.

For the velocity distribution of disc stars we adopt the results of \citet{Bond2010}. They model 
the distribution of rotational velocities as a sum of two Gaussians, first normalized to the factor of $0.75$ and second to $0.25$.
At the disc plane ($z = 0 \rm \, pc$) 
mean value and dispersion of the first Gaussian are $\mu_{\phi} = -194 \; \rm km \; s^{-1}$ and $\sigma_{\phi}
= 12 \; \rm km \; s^{-1}$ respectively, whereas the second Gaussian is given with $\mu_{\phi} = -228 \; \rm km \; s^{-1}$ 
and $\sigma_{\phi} = 34 \; \rm km \; s^{-1}$. Distributions of radial and vertical to the disc velocities 
are described with a single Gaussian only, for which the mean value $= 0 \; \rm km \; s^{-1}$ in both cases. 
Dispersions at the disc plane are given as $\sigma_{r} = 40 \; \rm km \; s^{-1}$ and 
$\sigma_{z} = 25 \; \rm km \; s^{-1}$. For all the distributions the values of dispersions increase slowly 
with distance from the disc $|z|$. 

Each binary is moving through the Galaxy subject to its Galactic potential. We have made use of the Galactic 
potential description of \citet{Flynn1996}, where the total potential is modeled by the sum of potentials 
of three standard components: the dark halo $\Phi_{H}$, the central component $\Phi_{C}$ and the disc 
$\Phi_{D}$. The dark halo potential is given in a spherical form:
\begin{equation}
 \Phi_{H} \propto V_H^2 \, {\rm ln}(r^2 + r_0^2)
\end{equation}
where $r^2 = x^2 + y^2 + x^2$ and $r_0 = 8.5 \rm \; kpc$, $V_H = 220 \rm \; km \; s^{-1}$. The spherical component potential is modeled 
by two separate components:
\begin{equation}
 \Phi_{C} \propto - \frac{M_{C_1}}{\sqrt{r^2 + r_{C_1}^2}} - \frac{M_{C_2}}{\sqrt{r^2 + r_{C_2}^2}}
\end{equation}
where $M_{C_1} = 3 \times 10^9 \; \rm M_{\odot}$ is the mass of bulge/stellar-halo, $M_{C_2} = 1.6 \times 10^{10} \; \rm M_{\odot}$ is the mass of the most central component, 
$r_{C_1} = 2.7 \rm \; kpc$ and $r_{C_2} = 0.42 \rm \; kpc$. The disc potential, on the other hand, 
is modeled by the sum of three Miyamoto-Nagai potentials \citep{Miyamoto1975}:
\begin{equation}
 \Phi_{D} = \Phi_{D_1} + \Phi_{D_2} + \Phi_{D_3}
\end{equation}
where:
\begin{equation}
 \Phi_{D_n} \propto - \frac{M_{D_n}}{\sqrt{\rho^2 + (a_n + \sqrt{z^2 + b^2})^2}} \; \; \; \; \; \; \; \; \; n = 1,2,3
\end{equation}
where $\rho^2 = x^2 + y^2$, mass-like parameters of subsequent disc components are $M_{D_1} = 6.6 \times 10^{10} \; \rm M_{\odot}$, 
$M_{D_2} = - 2.9 \times 10^{10} \; \rm M_{\odot}$, $M_{D_3} = 3.3 \times 10^9 \; \rm M_{\odot}$ and the values 
of other parameters are $a_1 = 5.81 \; \rm kpc$, $a_2 = 17.43 \; \rm kpc$, $a_3 = 34.86 \; \rm kpc$ and $b = 0.3 \; \rm kpc$.

We evolve each binary's position and velocity in the Galactic potential implementing Verlet's integration algorithm \citep{Verlet1967}, 
applicable for Newtonian equations of motion (often referred to as Velocity Verlet algorithm).

\subsubsection{Encounters with fly-bys}
\label{sec:encounters_evolution}

We model random encounters with passing-by stars utilizing the mean-free time $t_{\rm MF}$ quantity introduced in 
Section~\ref{sec:dynamic_theory}. We apply adaptive time step $\delta t$, dependent of the expected frequency of interactions 
for a given binary, limited to not exceed 1000 yr to guarantee smooth dynamics evolution:
\begin{equation}
\delta t  = {\rm min}(0.1 \; t_{\rm MF} \; \; ; \; 1000 {\rm yr})
\end{equation}
Every hundred time steps of a binary dynamics evolution the value of
$t_{\rm MF}$ is updated to correspond to the changing local conditions (\textit{i.e.} stellar number density and 
velocity distribution). It is also updated after each dynamical interaction.

Upon each time step $\delta t$ we randomize whether or not 
a flying-by field star was encountered during this time, with the probability of it equal to $\delta t / t_{MF}$. If a star was encountered, 
we draw three main parameters describing such event -- encountered star mass $M_{\rm single}$, its relative velocity to the binary
$\vec{v_{\rm rel}}$ and impact parameter $b$ -- following procedure described in Section~\ref{sec:dynamic_theory}. 
Subsequently, we employ \textsc{fewbody} code \citep{Fregeau2004} to simulate the interaction and obtain resulting system parameters. \textsc{fewbody}
uses the 8th-order Runge-Kutta Prince-Dormand integration method with adaptive time step to solve N-body problems for small, few body
systems (N < 10). It offers a suitable toolkit for binary-single scattering experiments -- it numerically integrates
orbits of stars during their interaction, terminates the computations once the result is unambiguous and classifies the resulting
configuration.
Apart from $M_{\rm single}$, $\vec{v_{\rm rel}}$ and $b$ each encounter is described by four additional parameters, 
which we draw from a flat distribution: mean anomaly $\xi$
of the binary and three angles defining the direction of a flying-by star movement with respect to the system's orbit. 

Once an encounter is completed we check whether or not the resulting orbit satisfies the condition for tidal circularization, i.e. 
whether $d_{\rm per} < d_{\rm tidal}$. If it does we keep monitoring its state until periastron is reached (i.e. $\xi = 0$.) 
which for very wide and eccentric orbits is roughly where tidal forces become significant.
For a wide binary of $a \sim 10^4 \, \rm AU$ it is possible to encounter third bodies at frequency comparable to its orbital 
period -- thus, the fulfilled condition for tidal circularization might not remain satisfied up until the periastron is reached 
as subsequent interactions could potentially violate it. 

\subsubsection{Tidal circularization}
\label{sec:tidal_circularization}

Eventually, once the periastron is reached and $d_{\rm per} < d_{\rm tidal}$ remains satisfied, the evolution of the 
binary's orbit becomes very complicated, as the amount of energy dissipated during a single periastron passage 
depends on tidally induced stellar oscillations and is very much a matter of debate.
If the timescale for dissipation of such oscillations' energy is short and a significant fraction of energy is
thermalized during one orbital period \citep[as argued by][]{Kumar1996},
the orbit becomes circularized quickly, with its size being reduced considerably already at the first pericenter 
passage. Such scenario is also supported by the results of \citet{Kaib2014}, who investigate periastron 
passages of wide, eccentric binaries in the regime of \citet{Press1977} and \citet{Lee1986} tidal modeling, 
and show that for pericenter separation of the order of 5 stellar radii the tidal circularization proceeds 
quickly, with pericenter distance remaining nearly constant (\textit{i.e.} $a \simeq d_{\rm per}$)

This simplified picture of tidal circularization becomes very complicated when we take into consideration 
that energy dissipated from tidally induced oscillations affects the internal structure of a tidally affected star \citep{Ray1987}. 
If the timescale of such process is too short, the heated star may expand and create a common envelope or even lead to 
a merger \citep{Podsiadlowski1996}. If, on the other hand, the energy dissipation is ineffective, the orbit may become subject 
to quasi-periodic fluctuations, significantly extending the circularization process \citep{Mardling1995}. 
In this work, however, we assume \citep[similarly to][]{Voss2007} that all binaries for which the pericenter is reached 
and $d_{\rm per} < d_{\rm tidal}$ is satisfied become circularized within one orbital period and $a \simeq d_{\rm per}$.

It should be noted here, that due to dynamic interactions some wide binaries can be excited into extremely wide orbits of $a \sim 10^{5} \rm AU$ 
and eventually disrupted, resulting in an observable plateau in $f(a)$ distribution \citep{Poveda2007}.
Such wide binaries are subject to exceptionally frequent stellar encounters, thus require a challenging amount 
of computational time. In order to address this issue we place an upper limit for a binary size $a_{\rm max} = 10^5 \rm AU$, 
at which we consider it certain to be disrupted and do not evolve it any further.
In the discussion we show that this limitation does not effect our predictions (see Sec.~\ref{sec:ultra_wide_section}). 

Before we move on to discuss the results we should clearly state an important simplification 
we make: similarly to MP2016 and other authors, we assume that {\em all} tidally circularized systems
become X-ray binaries (LMXBs precisely, as they all turn out to have low-mass companions). 
In reality, the evolution of a tidally circularized 
binary may not always lead to a mass transfer and X-ray activity. This is not only due to the uncertainties concerning the
influence of tidal heating on the MS stellar structure (as discussed in details earlier in this section), 
but also due to the fact that even after circularization, once the binary components are very close to each other, 
the MS secondary may not fill its Roche Lobe just yet and it may take even several Gyr for a mass 
transfer to be launched \citep[an exemplary time period obtained by][for a system of $\sim$ 7-10 $\rm M_{\odot}$ BH and $\sim$ 1--1.2 $\rm M_{\odot}$
secondary MS, reducing separation from 8 $\rm R_{\odot}$ to 5 $\rm R_{\odot}$ and initializing mass transfer, is $\sim$ 4 Gyr ]{Wiktorowicz2014}.
However, our typical separations after circularization are rather small
($a \approx 3-7 \rm \; R_{\odot}$ for nearly all the of systems in $\alpha = 0.5$ and $\alpha = 0.8$ models
and $a \approx 3-10 \rm \; R_{\odot}$ in the majority of cases in other models) 
and we expect that combined effects of magnetic braking and gravitational waves emission
will allow for a significant orbital decay in a lifetime of a binary leading to LMXB phase in majority of our systems. 

\section{Results}
\label{sec:results}

For the scope of this section we will denote the moment just after the BH formation as Zero Age (ZA) for BH-MS 
binaries. Out of $\rm 4.5 \times 10^7$ systems at the beginning of pre-BH stage 
about $\sim 27.5\%$ survive until ZA for BH-MS, with only $\sim 1.12\%$ of those having companions expected to remain on MS until
present day. Thus, the ZA BH-MS population of interest consists of roughly 138,500 systems, with the majority of secondaries 
being low-mass ($\rm < 2 M_{\odot}$) MS stars due to their longer 
living expectancy. A typical semi-major axis for ZA BH-MS population lies in range 100--1000 AU, 
whereas the distribution of eccentricities covers full range 0--1, peaking at $e \sim 0.1$.

\subsection{An exemplary evolutionary route}
\label{sec:ex_ev_route}

In order to illustrate characteristic features of evolution leading to dynamical formation of LMXBs we take a closer look 
at the evolutionary path of an exemplary binary which have become an LMXB out of a Galactic disk BH-MS binary 
through dynamics and tidal circularization. Note that this is the very same binary that served as an example when describing the details 
of typical pre-BH evolution in Sec.~\ref{sec:startrack}. Here we continue its journey towards tidal circularization, starting from 
a newly formed ZA BH-MS system comprising of a $\sim \rm 7.9 \, M_{\odot}$ BH and a $\sim \rm 0.75 \, M_{\odot}$ MS companion,
having orbital parameters $\rm a = 306 \, AU$ and $\rm e = 0.47$. 
The system has been assigned a Galactic velocity $\vec{v}$, drawn from the local stellar velocity distribution, 
which with respect to the mean local velocity $\vec{v_*}$ was $|\vec{v_{\rm rel}}| = |\vec{v_*} - \vec{v}| = \rm 75 \, km/s $. 
Since the BH was created with no direct natal kick and the Blaauw kick was very small, the system's velocity deviation $\vec{v_{\rm rel}}$ is 
only caused by dispersion of the Galactic velocity distribution. 

Then, during a $\sim$ 7.6 Gyr long period of evolution stretching from the BH formation to the binary tidal circularization, the system 
was subject to around 5,000 stellar encounters, affecting its orbital parameters and eventually leading to very high eccentricity 
and very small distance between companions at periastron. At the moment of tidal circularization the binary's orbital 
parameters were $a \approx \rm 1730 \, AU$ and $e \approx \rm 0.99999$, corresponding to separation at pericenter 
$d_{\rm per} \approx \rm 3.8 \, R_{\odot}$ and companion filling its Roche Lobe in 89$\%$ (radius-wise), which is 
sufficient for circularization in all our models.

Figure~\ref{fig:detailed_evolution} shows detailed evolution of the binary's
orbital parameters $a$ and $e$, as well as the pericenter separation $d_{\rm per}$. 
\begin{figure}
	\includegraphics[width=\columnwidth]{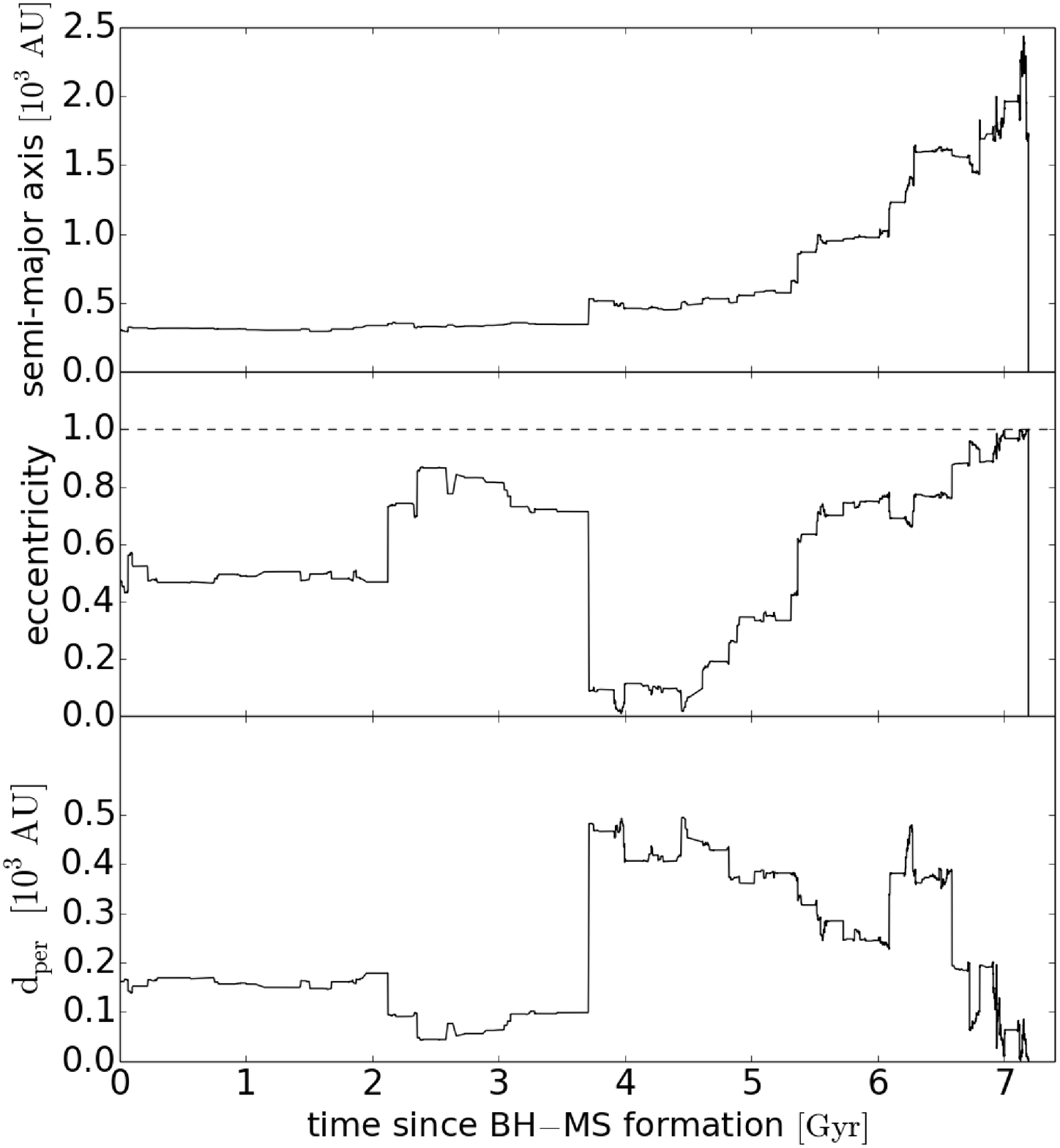}
    \caption{Detailed evolution of orbital parameters $a$ and $e$, as well as distance at pericenter $\rm d_{per} = a (1-e)$ for an exemplary, 
    typical binary that have undergone tidal circularization 
    (noticeable as a rapid decrease of both semi-major axis and eccentricity at the very end of evolution). This particular system was subject to 
    over 5,000 stellar encounters, most of them occurring during second half of its evolution. The binary was subject to several 
    major changes of orbital parameters due to encounters with exceptionally massive fly-bys. For instance, the most significant change of eccentricity 
    from $0.71$ to $0.09$ at around 3.7 Gyr (coupled with semi-major axis increase from $345$ AU to $530$ AU) was caused by an interaction with a
    $\sim 4.5 \rm \; M_{\odot}$ star, whereas the mean mass of fly-bys in our simulations is $\sim 0.62 \rm \; M_{\odot}$. Additionally, the encountered star
    was passing by with a relatively low speed of $\sim 45 \rm \; km \; s^{-1}$ with respect to the binary (the mean relative speed of all fly-bys
    being $\sim 125 \rm \; km \; s^{-1}$), which granted an exceptionally long time of gravitational interaction.}
    \label{fig:detailed_evolution}
\end{figure}
Even though the binary was affected by more than 5,000 encounters altogether the vast majority of interactions had very little effect on its orbit, with 
only few inducing significant changes of orbital parameters. 
\begin{table*}
	\centering
	\begin{tabular}{l | c | c | c | c | c | c} 
		\hline
		model & \multicolumn{2}{c|}{~~~~~$\rm {|\bar{v} -\bar{v}_*|}$ $\rm [km/s]$~~~~~~~~~~~~~~~~~~~~} 
		      & \multicolumn{2}{c|}{~~~~~~~~~~$\rm semi-major \; axis \; [10^3 AU]$~~~~~~~~~~~~~~~~~} 
		      & \multicolumn{2}{c|}{~~~~~~~~~$\rm eccentricity$~~~~~~~~~~~~~~} \\
		& ZA & CIRC & ZA & CIRC & ZA & CIRC\\
		\hline
		MP2016 & $68 \pm\; ^{42}_{27}$ & $73 \pm\; ^{43}_{32}$ & $0.42 \pm\; ^{0.39}_{0.23}$ & $1.38 \pm\; ^{7.32}_{1.0}$ & $0.32 \pm\; ^{0.38}_{0.21}$ & $\gtrsim 0.9999$\\[1ex]
		0.2 Roche & $67 \pm\; ^{40}_{26}$ & $74 \pm\; ^{42}_{33}$ & $0.42 \pm\; ^{0.35}_{0.23}$ & $1.41 \pm\; ^{7.19}_{1.01}$ & $0.32 \pm\; ^{0.39}_{0.21}$ & $\gtrsim 0.9999$\\[1ex]
		0.5 Roche & $71 \pm\; ^{37}_{28}$ & $78 \pm\; ^{44}_{36}$& $0.41 \pm\; ^{0.35}_{0.21}$ & $1.35 \pm\; ^{6.41}_{1.03}$ & $0.34 \pm\; ^{0.38}_{0.24}$ & $\gtrsim 0.9999$\\[1ex]
		0.8 Roche & $70 \pm\; ^{39}_{26}$ & $77 \pm\; ^{32}_{33}$& $0.40 \pm\; ^{0.35}_{0.20}$ & $1.14 \pm\; ^{3.69}_{0.75}$ & $0.38 \pm\; ^{0.34}_{0.28}$ & $\gtrsim 0.9999$\\[1ex]
		\hline
		all Zero Age & $46 \pm\; ^{29}_{21}$ & - & $0.28 \pm\; ^{0.35}_{0.17}$ & - & $0.20 \pm\; ^{0.34}_{0.11}$ & - \\
		BH-MS binaries\\
		\hline
	\end{tabular}
	\caption{The comparison of median values of characteristic binary parameters between the Zero Age (ZA) and the moment 
	of tidal circularization (CIRC), drawn for the population of binaries that have undergone circularization. 
	Also compared with median values of these parameters for the whole ZA BH-MS population. Subsequent columns correspond to 
	median values of: deviation of a binary velocity $\bar{v}$ from mean local stellar velocity $\bar{v}_*$, semi-major axis $a$ 
	and eccentricity $e$. Ranges corresponding to $68 \%$ of all systems (i.e. $1\sigma$) are given for each quantity except 
	the eccentricity at the moment of tidal circularization (rightmost column), for which all the systems satisfy $e\gtrsim 0.9999$. }
	\label{tab:second_table}
\end{table*}
The average mass of passing-by stars in the only 10 encounters during which the orbital energy was changed by more than 10 \% was 
$\sim \rm 13.5 \, M_{\odot}$, whereas the average mass of all stars encountered by the binary was $\sim \rm 0.62 \, M_{\odot}$. 
In fact, it is very typical for the majority of dynamically formed LMXBs in our simulations that the dynamic evolution of a binary is mainly dependent on a 
few key encounters with, on average, exceptionally massive fly-bys. This is connected to the fact that, on average, the amplitude of a velocity kick 
induced by a fly-by to one binary component with respect to the other can be analytically estimated as proportional to the fly-by's mass \citep[e.g.][]{Hills1981}, i.e.:
\begin{equation}
\label{eq:Hills1981}
\big< \Delta V \big> = \rm \frac{3 a G M_{\rm fly-by}}{v_{rel}b^2}
\end{equation}
where $\rm a$ is semi-major axis of the binary, $\rm M_{fly-by}$ is mass of the fly-by, $\rm b$ is the impact parameter and $\rm v_{rel}$ is the fly-by's velocity, relative to mass center of the binary.
This suggests that the local stellar mass distribution has influence on the timescale of evolution through dynamical interactions and
therefore the effectiveness of LMXBs' dynamical formation channel. We discuss it further in Sec.~\ref{sec:mass_func_flybys}.

\subsection{Evolution of orbital parameters}
\label{sec:evol_orb_param}

The evolution of orbital parameters at figure~\ref{fig:detailed_evolution} also illustrates a general trend of steadily increasing semi-major axis. This 
is an average regularity for all of our evolutionary paths leading towards tidal circularization, which
is shown in table~\ref{tab:second_table}, where we compare some basic characteristics of binaries which have evolved to undergo a circularization
with same characteristics of the whole BH-MS population. We choose parameters which are influenced by stellar iterations: 
the deviation of a binary velocity $\bar{v}$ 
from mean local stellar velocity $\bar{v}_*$, as well as the values of orbital parameters $a$ and $e$. We analyze those 
parameters at two moments of time: at ZA for BH-MS systems (i.e. just after the BH formation) and at the beginning of tidal circularization (CIRC). 

Due to stellar encounters the semi-major axes of future BH LMXBs tend to be gradually increasing, with their typical values at the moment
of circularization being a few times higher than at ZA. This, paired with extremely high eccentricities induced by encountered fly-bys ($e \gtrsim 0.9999$),
makes the orbit of a typical binary at circularization very wide and almost linear (i.e. degenerate ellipse). Such increase of a semi-major axis, 
very characteristic for our population of systems to undergo tidal circularization, makes a binary subject 
to more frequent stellar encounters as wider systems have naturally larger cross-sections (see relation~\ref{eq:NOI} for which $b_{\rm max} \propto a$).
This also explains the slight bias towards wider ZA BH-MS systems in the population of future BH LMXBs -- for each of our models 
the majority of systems to undergo tidal circularization had ZA separations falling in range $200-750 \rm \; AU$, whereas 
the corresponding most common range for the entire BH-MS population was around $100-600$ AU.

According to relation~\ref{eq:NOI} systems with higher velocity deviations from typical field stars velocities are 
subject to more frequent interactions with fly-bys. Our results show that they are thus more predisposed to be tidally circularized, 
as the typical values of velocity deviations for future LMXBs (falling in range $\sim 40-110 \; \rm km \; s^{-1}$) are significantly higher then 
velocity deviations for the entire BH-MS population ($\sim 25-75 \; \rm km \; s^{-1}$). However, only a negligible $< 1\%$ of all our BH LMXB systems
have been given a non-zero natal kick during BH formation -- their distinctly high velocities have been drawn from the Galactic velocity 
distribution due to its dispersion.
In general, the effect that potentially higher BH natal kicks could have on the effectiveness of the dynamical formation channel is unclear.
On the one hand higher kicks could potentially lead to formation of more BH-MS systems moving through the Galaxy with exceptionally 
high velocities, likely subjects of numerous stellar encounters. On the other hand, stronger kicks could significantly lower the number of binary 
systems surviving BH formation in the first place. 
We further discuss the later effect in Sec.~\ref{sec:bh_natal_kicks}.

\subsection{The population of LMXBs}

Table~\ref{tab:first_table} compares the numbers of LMXBs ($\rm N_{LMXB}$) formed dynamically in 4 tested models for tidal circularization 
threshold distance, as well as the values
of donor mass $\rm M_{MS}$, also compared with the whole ZA BH-MS population,
the mean times between ZA and circularization ($\rm t_{evol}$, which is roughly mean age of circularized binaries) and finally 
the mean numbers of interactions (NOI) which had occurred by the time circularization was achieved.

The amount of dynamically formed LMXBs decreases for different models from $\sim 220$ for the MP2016 condition for circularization 
down to $\sim 65$ in a model, where circularization occurs only when secondary fills at least 80 \% of its Roche Lobe (radius-wise).
This corresponds to the fact that for subsequent models the conditions for circularization becomes more and more 
stringent for low values of mass ratios $q \lesssim 0.25$ expected in BH-low mass MS binaries -- see the comparison of $d_{\rm tidal}$ vs mass
ratio $q$ relations at figure~\ref{fig:d_tidal_models}.

\begin{table*}
	\centering
	\begin{tabular}{l | c | c | c | c} 
		\hline
		model & ~~~~~~~~~~$\rm N_{LMXB}$~~~~~~~~~ & ~~~~~~ ~~$\rm M_{MS}$~~~~~~~~ & ~~~~~~~~~~~~$\rm t_{evol}$~~~~~~~~~~~~ & ~~$\rm NOI$~~~\\
		& & $\rm [M_{\odot}]$ & $\rm [Gyr]$ & $[10^3]$\\
		\hline
		MP2016 & $219$ & $0.64 \pm\; ^{0.44}_{0.42}$ & $1.6 \pm\; ^{2.9}_{1.3}$ & $1.5 \pm\; ^{4.3}_{1.1}$\\[1ex]
		0.2 Roche & $189$  & $0.81 \pm\; ^{0.44}_{0.50}$ & $1.5 \pm\; ^{3.0}_{1.3}$ & $1.6 \pm\; ^{5.3}_{1.1}$\\ [1ex]
		0.5 Roche & $104$ & $0.85 \pm\; ^{0.42}_{0.47}$ & $1.4 \pm\; ^{3.4}_{1.1}$ & $1.5 \pm\; ^{6.2}_{1.0}$\\[1ex] 
		0.8 Roche & $64$ & $0.81 \pm\; ^{0.49}_{0.44}$ & $1.6 \pm\; ^{4.3}_{1.2}$ & $1.6 \pm\; ^{7.0}_{1.0}$\\[1ex]
		\hline
		all Zero Age & - & 0.81 $\pm\; ^{0.83}_{0.51}$  & - & -\\
		BH-MS binaries\\
		\hline
	\end{tabular}
	\caption{Comparison of dynamically formed LMXBs for different tidal circularization 
	models. Subsequent columns correspond to: the number of formed LMXBs ($\rm N_{LMXB}$) and the median values of the mass of MS companion $\rm M_{MS}$, 
	the time between BH-MS formation and circularization ($\rm t_{evol}$) and the number of interactions ($\rm NOI$) which occured
	before circularization. Ranges corresponding to $68 \%$ of all systems (i.e. $1\sigma$) are given. Subsequent 
	rows correspond to different models. The very final row corresponds to the entire BH-MS population at ZA. }
	\label{tab:first_table}
\end{table*}
\begin{figure}
	\includegraphics[width=230px]{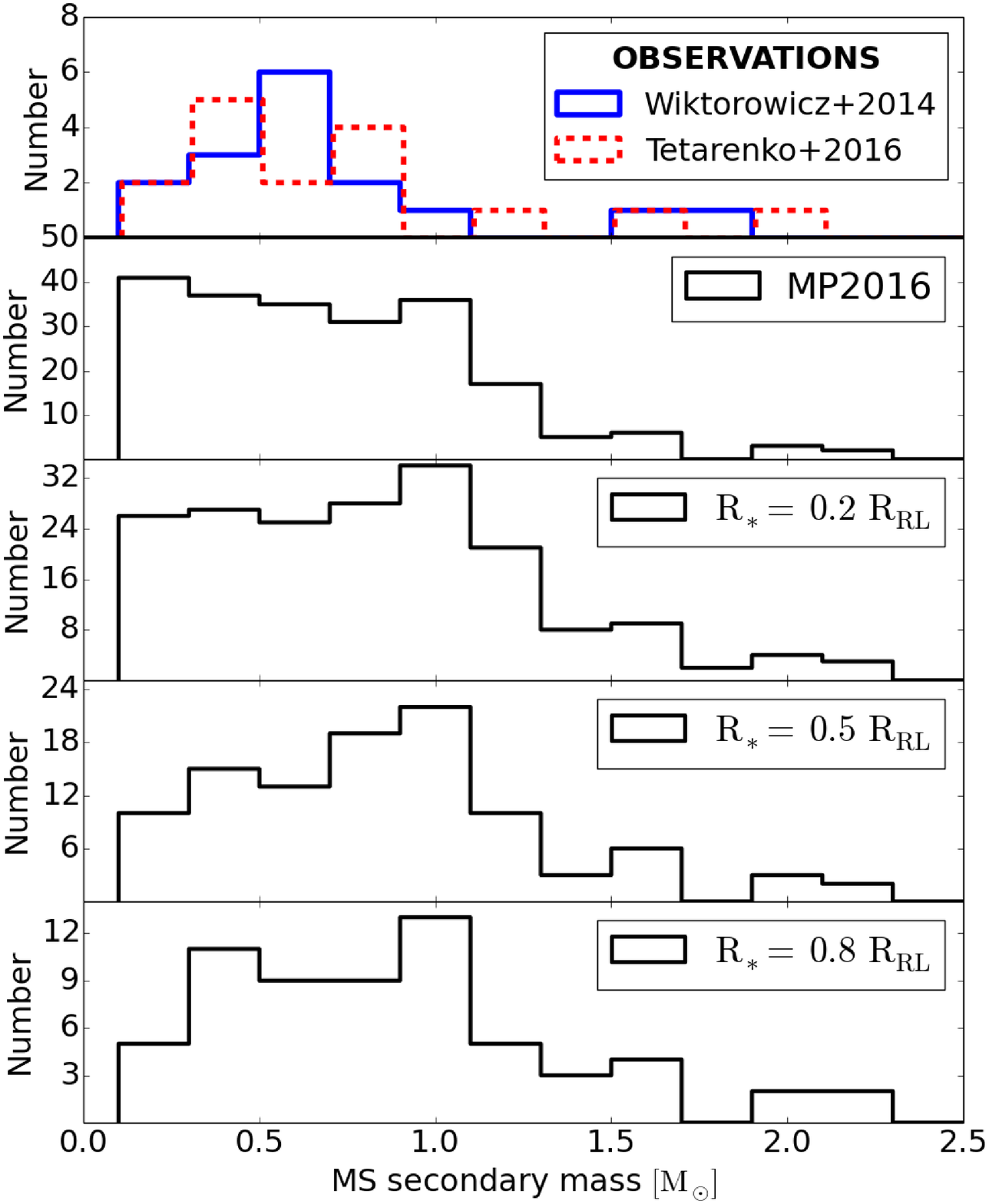}
    \caption{The comparison of companion mass function of tidally circularized systems for all our models 
    as well as the observed population of Galactic BH LMXB systems (data taken from table 1 of \citet{Wiktorowicz2014}, 
    alternative values come from table 13 of \citet{Tetarenko2016B}). 
    Shape of each distribution is strictly connected to the shape of $d_{\rm tidal}(q)$ relation in a given model, favoring less massive companions for models with 
    significantly higher $d_{\rm tidal}$ values for low mass ratios. It is worth noticing that in all models the majority of companions lie within the 
    range 0-1 $\rm M_{\odot}$, which is consistent with observations. However, a robust comparison 
    would require modeling of the mass transfer phase (see text).}
    \label{fig:mass_distribution}
\end{figure}

The differences between $d_{\rm tidal}(q)$ relation for subsequent models are also reflected in corresponding differences
between the shape of companion mass function (CMF) of a resulting population of dynamically formed BH LMXBs (see figure~\ref{fig:mass_distribution}).
For MP2016 and 0.2 Roche Lobe models, where the values of $d_{\rm  tidal}$ 
are significantly higher for low mass ratios, so that circularization becomes easier for less massive stars, 
the shape of CMF responds respectively. 

For comparison we also plot the CMF of observed Galactic BH LMXB systems. One set of data was taken from table 1 of \citet{Wiktorowicz2014}
whereas alternative values come from table 13 of \citet{Tetarenko2016B} (the differences mainly arise from that fact that \citet{Tetarenko2016B}
applies BH masses argued by \citet{Kreidberg2012}).
Most of the companions with measured masses are found in range 0.1--1.0 $\rm M_{\odot}$, which is consistent with our findings. 
However, a more robust comparison with the observed distribution (i.e. one taking into account its shape for M < $\rm M_{\odot}$) is beyond the 
scope of this paper, as it would require modeling of the actual LMXB phase and mass transfer rates.
Also note that the sample of systems with measured companion masses is still very small ($\sim$20) and the 
masses themselves may be subject to systematic errors \citep{Kreidberg2012}. 

We wish to mention that the observed CMF is in tension with the results of population synthesis 
of LMXBs through CE evolution for various CE energy binding parameters proposed up to date \citep{Wiktorowicz2014}, 
as they generally result in a distribution peaking strongly around 1.0 $\rm M_{\odot}$ with a possible smaller peak around 0.2--0.3 $\rm M_{\odot}$. 
Similar results were obtained by \citet{Wang2016}, although it seems that a better agreement with observations could potentially be obtained if BH
were formed in failed SN \citep{Kochanek2014} rather than through material fallback \citep{Fryer2012}.
In the meantime channels involving intermediate mass donors \citep{Justham2006,Chen2006} are 
also inconsistent with observations as they predict donor spectral types 
which are too hot to explain the observed temperatures of companion stars in BH LMXBs.

\section{Discussion}
\label{sec:discussion}

Table~\ref{tab:first_table} shows that the number of LMXB candidates is strongly dependent on the assumption on pericenter 
distance required for tidal circularization, varying by a factor of $\approx 3.5$ across our models. In fact, even our most 
pessimistic model may be too optimistic, as strong tidal interactions at pericenter may not always lead to circularization 
of the orbit (see \citet{Podsiadlowski1996,Mardling1995} and the brief discussion in Sec.~\ref{sec:tidal_circularization}).

In the following text we point out main assumptions of our analysis and discuss their significance.

\subsection{Lifetime of LMXBs}

\label{sec:lifetimes}

Because the process of tidal circularization is still uncertain, both in terms of the timescale and the final outcome
(see Sec.~\ref{sec:tidal_circularization}), we do not 
attempt to model the evolution of our systems once they fulfill our condition for tidal circularization. Instead
we simply assume that each of such systems will have enough time to become an X-ray binary 
and it contributes to the final numbers of LMXB systems in table~\ref{tab:first_table}.

In particular we do not model the actual phase of X-ray activity and do not account for its finite duration.
This leads to an overestimation of the number of LMXBs in our predictions. 
For instance, if we assume that it typically takes up to 2 Gyr for a system to start its X-ray phase once 
it fulfills the condition for tidal circularization 
and that the duration of the following LMXB phase is about 1 Gyr 
we find that only about $20\%$ of the systems given in table~\ref{tab:first_table} can be observable as 
X-ray binaries at present time. 
Because the typical mass transfer rates from low-mass MS donors discussed in the literature are of the order 
of $\rm \sim 10^{-10} \, M_{\odot} \, yr^{-1}$ \citep{King1988,Verbunt1993}, which corresponds to LMXB lifetimes
of about several Gyr, the actual fraction of observable systems is likely higher -- somewhere between $20\%$ and $100\%$. 

\subsection{Mass function of the fly-bys}
\label{sec:mass_func_flybys}

The analyses of the evolution of orbital parameters due to interactions with passing-by stars (Sec.~\ref{sec:ex_ev_route})
reveals that the most significant changes are, on average, induced by massive fly-bys of at least several solar masses. 
This suggests that the local stellar mass function (SMF) has influence on the timescale of evolution through dynamical interactions and
therefore the effectiveness of LMXBs' dynamical formation channel. The SMF of the Milky Way disc is 
generally unknown, as there is no reliable way of constraining the population of Galactic dark objects such as
single black holes \citep[although there is some hope in the possibility of observing microlensing events][]{Wyrzykowski2016}.
In this study we simply derive masses of flying-by stars from the IMF of \citet{Kroupa1993} with the power-law exponent modified to $-2.3$
for the stars of mass $\rm M> 1.0 \, M_{\odot}$, as recently adjusted by \citet{Sana2012}. In reality though the SMF should be shifted 
towards lower masses compared to the IMF, since giant stars evolve and lose mass very quickly. Thus, even though 
we do not account for binary and higher order fly-bys, our approach still likely overestimates the frequency 
of interactions with massive systems.
For comparison, if we employ \textsc{StarTrack} for single star evolution, 
assume the above-mentioned IMF, constant star formation rate of 10 Gyr, 50\% binary fraction and 
that the average binary mass is $\rm \sim 1.5$ 
the average single star mass (coming from flat mass ratio distribution) we obtain the present SMF
which returns massive fly-bys (of at least 5 Solar masses) $\sim$ 3.5 times less often than the IMF we used. 
If we now assume that only encounters with such massive objects play any role in evolution 
of orbital parameters, in which case the times between BH-MS formation and circularization of our systems
would be 3.5 times longer, we find that only about $\sim$ 40\% of  
LMXB candidates in table~\ref{tab:first_table} would have enough time to become circularized 
(i.e. $N_{\rm LMXB}$ reduced by a factor of $\sim$ 2.5), irrespective of the model of tidal circularization. 
Such assumption is pessimistic since encounters with less massive fly-bys can also have an effect on the binary orbit 
given the impact parameter $\rm b$ is low (see equation~\ref{eq:Hills1981}).

\subsection{BH natal kicks}
\label{sec:bh_natal_kicks}

Because wide BH-MS binaries have relatively low orbital energy and can be easily disrupted, natal kicks play a crutial 
role in their formation. Unfortunately, the BH natal kicks are still poorly understood \citep[eg. ][]{Repetto2015}. They are generally 
thought to be caused by two main processes: asymmetries in the SN ejecta (when the BH is believed to be
formed via fallback onto the proto-NS, see \citealt{Fryer2001,Zhang2008}) 
or asymmetries in neutrino flux (also possible during direct collapse and BH formation without SN, 
although it might require strong magnetic fields -- see \citealt{Janka2013} for the recent review).
In either case the
assumption of linear momentum conservation yields BH natal kicks reduced with respect to those of NS: 
$\rm V_{BH \, NK} \sim (M_{NS}/M_{BH}) V_{NS \, NK}$. This simple picture implies BH natal kicks 
decreasing with BH mass, which is in agreement with the observations of Galactic BH X-ray binaries 
\citep[although natal kick velocities independent of BH mass cannot be ruled out yet][]{Belczynski2016}.
It might be challenged, however, by the spacial distribution of known systems in the Milky Way, 
which seems consistent with BH gaining similar velocities to NS rather than similar momentum \citep{Repetto2012}, 
although \citealt{Mandel2016} argues that with a more careful approach and within uncertainties no BH kicks 
larger than 100 km/s are required to explain all observed systems.
While natal kicks seem necessary for some binaries to reach the large distances 
above the Galactic plane at which they are observed, the positions of several systems 
can be explained with no natal kick whatsoever \citep{Repetto2015}. This could correspond 
to the case in which a BH is formed directly from a massive star with no associated SN, no mass loss
and no asymmetry in neutrino emission. 

We have employed asymmetric mass ejection model for BH natal kicks in
combination with rapid supernova model for BH mass \citep{Fryer2012}. 
This results in magnitude of the kick decreasing with BH mass (with no 
natal kicks for most massive BHs).
This is consistent with above-mentioned observations as well as with current 
LIGO limits on merger rates of BH-BH binaries \citep{Belczynski2016_nature}. 
However, we note that models with moderate BH natal kicks ($\sim 100-200$ km/s)
are \emph{also} consistent with all those observations and limits 
\citep{Belczynski2016}.
Because the wide BH-MS systems evolve from wide MS-MS binaries (a $\gtrsim$ 100 AU) there is no 
mass transfer between companions.
The BHs in these systems evolve from non-interacting stars following single
stellar evolution with BH mass set by the initial star mass, metallicity and 
our adopted wind mass loss and supernova model.
For \citet{Sana2012} IMF the rapid SN model results in 
about 56\% of BHs receiving zero natal kicks (direct collapse),
while for the rest 44\% the amount of material falling back is always bigger than 62\%  corresponding to 
$\rm \sigma_{NK} \lesssim 100 \, km/s$. Characteritic orbital velocity 
$\rm V_{orb} \sim \sqrt{G(M_a + M_b)/a}$ for our systems at the moment of BH formation 
($\rm M_a \sim 10 \, M_{\odot}$, $\rm M_b \sim 1 \, M_{\odot}$, $\rm a \sim 300 \, AU$)
is $\rm V_{orb} \sim 6 \, km/s$. This indicates that unless the BH natal kick is very small, 
such systems are likely to be disrupted (which indeed is the case for almost all of the 44\% of our BHs which 
do not form in direct collapse). In fact, only about 10\% and about 30\% of the BH LMXB 
candidates from table~\ref{tab:first_table} would survive the BH formation 
if NKs were drawn from a Maxwellian with $\sigma_{NK} = 10$ km/s or even only $\sigma_{NK} = 5$ km/s, 
respectively. Comparison of different BH NK models is beyond the scope of this study but 
it seems necessary that a significant fraction (in our case $\sim$56\%, direct BH formation)
of BHs is born with zero to negligible NK of a few km/s in order to produce 
a large enough population of wide BH-MS systems, so that some of them may be 
circularized thanks to stellar encounters and become LMXBs.

\subsection{Ultra-wide systems -- comparison with MP2016}

\label{sec:ultra_wide_section}

As discussed in Sec.~\ref{sec:startrack}, the multiplicity of systems with massive stars 
has been found to increase with separation \citep[eg.][]{Sana2014,Aldoretta2015}. In particular \citet{Moe2016}
reports that ZAMS companions to massive stars on orbits with periods
$\rm log(P/day) \gtrsim 5.5$ are essentially tertiary components in hierarchical \textit{triples}, 
which makes them unlikely progenitors of BH LMXBs through dynamical scenario tested in this study (see Sec. 3.1.4 of MP2016).
For that reason we set $\rm log(P/day) = 5.5$ as the upper limit on the initial binary periods,
which results in $a_{\rm MS-MS} \lesssim 10^{2.5}$ AU for the MS-MS binaries and
$a_{\rm BH-MS} \lesssim 10^{3}$ AU for BH-MS systems just after the BH formation. 
On the contrary, MP2016 only consider BH-MS binaries with $\rm 10^3 AU < a < 3 \times 10^4 AU$, 
for which they predict the dynamical scenario to be most likely. Thus, 
for the sake of comparison with their result, we have simulated a sample of additional $1.215 \times 10^7$ \textit{binaries}
with periods at ZAMS between $\rm log(P/day) = 5.5$ and $8.5$, which we denote as ''ultra-wide systems''
to differentiate from our main sample -- the ''wide systems''. For the rest of this section we also assume MP2016 
condition for tidal circularization. 

MP2016 predict that the probability of a dynamical interaction causing tidal capture 
and circularization is the highest for 
$a_{\rm crit} \approx 1.3 \times 10^4 $AU (see their Fig.1).
In order to compare this prediction with our simulations 
we combine BH LMXB candidates from both populations, wide and ultra-wide systems,
and plot their semi-major axes at the moment just before the final interaction, 
after which the MP2016 condition for tidal 
circularization is fulfilled (see figure~\ref{fig:ultra_wide_systems} top panel). 
We indeed find that the distribution has its maximum at around $10^4$ AU, which 
is in good agreement with MP2016 estimation. Note that our upper cut-off at $10^5$ AU on the binary 
separation above which we stop following its evolution (marked with vertical line in the plot) 
is at the falling end of the distribution, so only few BH LMXB candidates were lost by introducing it. 

\begin{figure}	
	\includegraphics[width=\columnwidth]{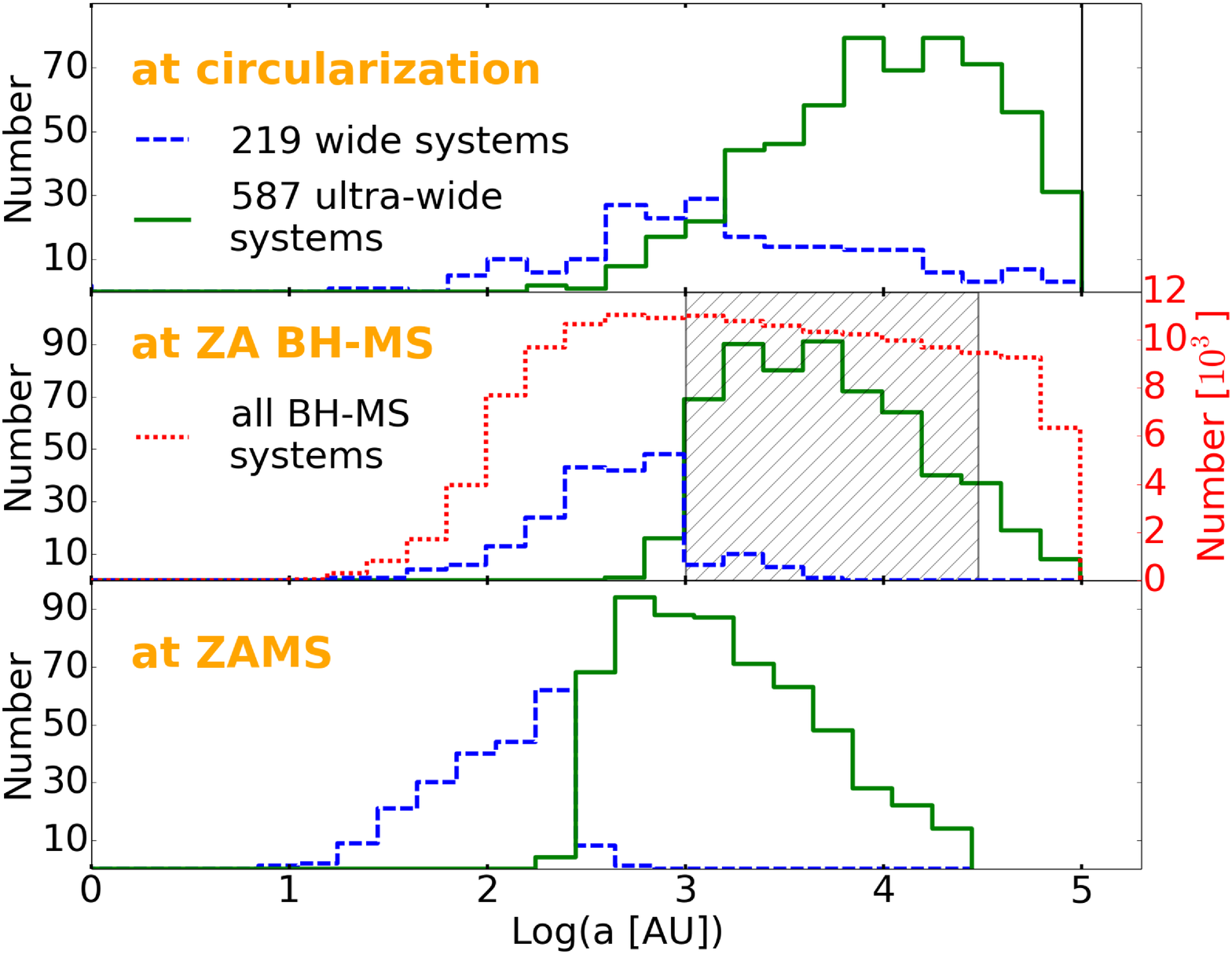}
    \caption{Distribution of semi-major axes for BH LMXB candidates from both wide and ultra-wide samples at three different 
    stages: at the moment just before the final interaction, 
    after which the MP2016 condition for tidal 
    circularization is fulfilled (top panel), at the moment of BH formation (middle panel) 
    and at ZAMS (bottom panel). 
    Wide systems are those which contribute to our main results (Sec.~\ref{sec:results}) while 
    the population of ultra-wide systems is an additional sample calculated to make a comparison 
    with MP2016. In the middle panel we also plot the semi-major axes distribution of 
    all our BH-MS systems, from both wide and ultra-wide samples (note the separate label on the right-hand side).
    Dashed region in the center panel marks limits on separation for BH-MS systems 
    assumed in MP2016. Vertical line at $a = 10^5$ AU in the upper panel corresponds to our limit 
    for a binary separation. The presence of maxima has been predicted by MP2016 and is in 
    agreement with their estimates (see text).
    }
    \label{fig:ultra_wide_systems}
\end{figure}

In the figure~\ref{fig:ultra_wide_systems} we also plot the distributions of semi-major axes 
at previous stages of the systems evolution: at ZA for BH-MS binaries and finally at ZAMS.
Since both the stellar evolution of BH progenitor (see Sec.~\ref{sec:startrack}) and dynamical interactions during the BH-MS phase
(see Sec.~\ref{sec:evol_orb_param}) tend to increase the binary semi-major axes, subsequent distributions in the plot are more and more 
shifted towards lower values. 
Note that the entire range of BH-MS separations $\rm 10^3 AU < a < 3 \times 10^4 AU$ considered 
by MP2016 (marked as dashed region) is covered. In the middle panel we also 
plot the semi-major axes distribution of all our BH-MS systems, from both wide and ultra-wide samples 
(the red, dotted curve with separate label on the right-hand side). 
In range from $10^{2.5}$ to $10^{4.5}$ AU it turns out to be relatively flat. 
This means that the MP2016 dynamical scenario is most effective 
for BH-MS systems of separations between 2000 and 6000 AU, corresponding to the maximum of the
distribution of BH LMXB candidate separations at ZA for BH-MS stage. 

For the MP2016 model of tidal circularization there are 587 BH LMXB candidates
in our sample of ultra-wide binaries and 219 among wide systems, giving 806 in total. 
Thus, the addition of ultra-wide systems increased predictions by a factor of about $\sim3.7$
(note that this factor is roughly the same for other models as well).
In order to compare these numbers with MP2016 results we need to adjust a few assumptions. 
Firstly, MP2016 only consider BH-MS systems with separation between $10^3$ and $3 \times 10^4$ AU, 
which results in 546 BH LMXB candidates from our samples. They also assume an LMXB lifetime of 1 Gyr, which reduces this number down to 138 binaries,
as well as a distribution of initial mass ratios $f(q) \sim q^{-0.5}$ \citep{Duchene2013},
which is shifted towards lower $q$ with respect to the flat distribution we adopt. 
If we assume that BH LMXBs only originate from systems with $q_{\rm ZAMS} \lesssim 0.05$  (masses of BH progenitors are over
20 $M_{\odot}$ at ZAMS, while MS companions are mostly low-mass stars of $M< 1 M_{\odot}$) this gives us scaling factor 
of $\sim 4.5$ and finally we obtain 621 BH LMXB candidates for MP2016 conditions. The only remaining difference 
is in the BH natal kicks. MP2016 estimate the number of BH LMXBs to be $N_{\rm LMXB} \approx 1200$ assuming no NKs for BH
and only $N_{\rm LMXB} \approx 60$ for NKs drawn from $\sigma_{BH;NK} = 19$ km/s (see their Table 3).
As we discuss in Sec.~\ref{sec:bh_natal_kicks} our model is somewhere in between ($56$\% of BHs receiving no NK),
and we expect that if $100$\% of our BHs received no NK the number of BH LMXB candidates from our simulations would be around 1100, 
which is in very good agreement with MP2016 prediction for that favorable case (around $1200$ systems). 
However, we would like to emphasis that this number is also based on three other optimistic assumptions: 
(i) the MP2016 condition for tidal circularization (more conservative models we tested give around $2$--$3.5$ times less BH LMXB candidates),
(ii) the assumption that mass transfer will be launched in all tidally circularized systems, leading to LMXB phase in each case (this might not always be the case, see \citealt{Ray1987} 
and Sec.~\ref{sec:tidal_circularization})
and finally (iii) the assumption that primordial binaries with massive BH progenitors are also born on ultra-wide orbits with periods of up to 
$\rm log(P/day) = 8.5$, while observations suggest that most of systems this wide are in fact hierarchical triples \citep{Moe2016}. 
A more conservative limit on initial period $\rm log(P/day) = 5.5$ results in about $3.7$ less BH LMXB candidates. 

We wish to mention that formation channels of LMXBs from triple star systems, 
although beyond the scope of this study, were also proposed. 
Most notably, orbits of inner binaries in hierarchical triples in which the outer companion is on a highly inclined 
orbit could be driven into high eccentricities through the eccentric Lidov-Kozai effect (see \citealt{Naoz2016b} for the recent review)
and form LMXBs this way. Recently \citet{Naoz2016} 
It was also suggested that exotic binaries such as LMXBs could originate from dynamically destabilized triples \citep{Perets2012b}.

On a final note, notice that some ultra-wide BH-MS progenitors face the danger of being disrupted during their pre-BH 
evolution due to extensive wind mass-loss. In the \citet{Vink2011} model of stellar winds a massive star ($\gtrsim$ 22 $\rm M_{\odot}$)
with Solar metallicity loses more than half its mass in just a fraction of a Myr during the core helium burning phase.
This means that for systems with low-mass secondaries ($\lesssim$ 2 $\rm M_{\odot}$) and semi-major axes of more than about
$10^4$ AU the binary loses over half its mass in less than one orbital period, which may lead to disruption.

\section{Conclusions}
We have performed the first population synthesis study on dynamical formation of BH LMXBs from primordial binaries in the Galactic field. 
Starting from a population of $4.5 \times 10^7$ primordial binaries with massive ($> 19 \rm M_{\odot}$) primaries of periods of up to $\rm log(P/day) = 5.5$, 
we utilize the \textsc{StarTrack} 
binary population synthesis code to produce a population of BH-MS binaries. Subsequently, we distribute them in the Milky Way disc following 
stellar density distribution and evolve their positions in the Galactic potential. Throughout this long evolutionary stage we 
account for random encounters with field stars and simulate results of such dynamical interaction using \textsc{fewbody} toolkit. We model 
tidal circularization of highly eccentric systems at periastron by applying a threshold distance $d_{\rm tidal}$, at which tidal 
forces become significant and are able to fully circularize the system. We test four models, each with a different prescription for
the value of $d_{\rm tidal}$: one suggested by MP2016 in their analytical approach and three more based on the ratio $\alpha$ of MS secondary radius 
to its Roche Lobe: $\alpha = 0.2$, $\alpha = 0.5$ and $\alpha = 0.8$ (Roche Lobe filling factor). 
Additionally,
we have evolved a 
sample of $1.215 \times 10^7$ ''ultra-wide'' binaries with initial periods between $\rm log(P/day) = 5.5$ and $8.5$ 
(corresponding to BH-MS systems forming with separations of about $10^3-10^5$ AU). 
Note, however, that systems this wide containing massive primaries 
are most likely triples \citep{Moe2016}.

We arrive at the following conclusions:
\begin{enumerate}
 
 \item \textbf{The number of BH LMXBs:} we find that dynamical formation of LMXBs from primordial binaries in the Galactic field
 can lead to a population of $\sim$ 60--220 BH LMXBs, 
 depending on the model of tidal circularization. 
 This number could possibly be higher by a factor of about 4.5 if 
 the distribution of initial mass ratios was $f(q) \sim q^{-0.5}$ \citep{Duchene2013},
 instead of the flat distribution we adopted.
 In the same time though, we do not account for the
 finite duration of the LMXB phase and we simplistically assume initial mass function for the mass function of flying-by field stars. 
 Both those assumption lead to an overestimation in the number of BH LMXB candidates, possibly be a factor of 5 and 2.5 respectively 
 (see Sec.~\ref{sec:lifetimes} and~\ref{sec:mass_func_flybys}).
 
 Our resulting population is most likely a small fraction of the expected total 
 number of BH LMXBs in the Milky Way, as the estimates based on 
 observational surveys typically span over the range of $10^2$--$10^4$ systems (e.g. \citealt{Romani1998}; 
 \citealt{CorralSantana2016}), whereas the recent discovery of the first candidate for quiescent BH LMXB outside of a globular cluster by \citet{Tetarenko2016}
 indicates, that quiescence BH LMXB may be even more abundant in the Galaxy than previously thought, rising the estimate of the total number of
 BH LMXBs up to $10^4$--$10^8$. Our results are based on a simple treatment of tidal circularization during sufficiently close
 periastron passages, where we assume that mass transfer will be launched in all tidally circularized systems, leading to a LMXB formation in 
 each case, which, in reality, might not always be the case \citep{Ray1987}. 
 Thus, our estimates serve more as upper limits and we consider it unlikely that a more detailed
 study could obtain a much bigger population of $\sim 10^4$ dynamically formed BH LMXBs. Our results rather indicate that the dynamical formation channel is only 
 responsible for a small fraction of all the Galactic sources -- unless the population of wide 
 BH-MS binaries is much larger than we considered. Possible additional sources of such systems include 
 ultra-wide primordial binaries with massive BH progenitors (with orbital periods beyond the limit $\rm log(P/day) = 5.5$ we assumed), 
 which could increase the number of dynamically formed BH LMXBs by a factor of about 3.5--4 
 (see Sec.~\ref{sec:ultra_wide_section}), as well as the cluster-dispersal scenario in which a massive object (such as a stellar BH)
 tidally captures a low-mass star on a wide orbit following a dispersal of their host cluster \citep{Perets2012}. 
 Another possibility for dynamical formation of LMXBs are channels involving triple star systems \citep[eg.][]{Perets2012b,Naoz2016}. 

 \item \textbf{Important factors :} 
  We find that wide ($ \gtrsim 200 \rm \; AU$), fast moving BH-MS binaries ($40-110 \; \rm km \; s^{-1}$ relative to local Galactic velocity) are most 
 likely candidates for dynamical formation of BH LMXBs. For such systems the frequency of stellar encounters is sufficient 
 (on average several thousand interactions during a few Gyr evolution), so that it is possible 
 for their orbits to be perturbated into highly eccentric states and then circularized by tidal forces at close pericenter passages. 
 In principle, the dynamical scenario would most efficient for systems with separations between $2000$ and $6000$, 
 although the existence of BH-MS binaries this wide is uncertain (see Sec.\ref{sec:ultra_wide_section}).
 Note that high velocities of our systems are solely associated with dispersion of the Galactic velocity distribution, as 
 the natal kicks of BHs in the surviving binaries in our simulations were either very small (a few km/s) or negligible. 
 In fact, we find that the dynamical scenario quickly becomes ineffective with increasing magnitude of BH natal kicks as only about 10\% of our
 BH LMXB candidates could survive BH formation with kicks drawn from an Maxwellian distribution with $\sigma_{NK} = 10$ km/s (see Sec.~\ref{sec:bh_natal_kicks}).
 Even though higher kick velocities could potentially produce fast-moving systems with respect to the surrounding stars, 
 giving chance for more stellar encounters and working in favor of the dynamical scenario, we consider this effect to be of secondary importance.
 
 We also show that the effectiveness of dynamical formation 
 channel is largely dependent on the effectiveness of tidal circularization, which is very much a matter of debate. 
 A detailed treatment of tidal forces would narrow the constraints on the number of formed BH LMXBs.
 
 We find that a typical evolutionary path leading towards a highly eccentric binary is dominated 
 by only a handful ($\sim$ 10) of truly significant interactions (i.e. orbital energy changing by more than 10 \%), even though the total number of encounters is 
 usually of the order of several thousands. 
 These few key interactions are, on average, with exceptionally massive fly-bys (often $>\rm 10 \; M_{\odot}$).
 
 \item \textbf{Companion mass function:} we find that the distribution of donor masses in dynamically formed BH LMXB systems is very sensitive 
 to changes of the condition for tidal circularization at close pericenter passage, which is still a poorly understood and difficult to model process. 
 Thus, a more detailed treatment of tidal forces is required to conclude about the exact shape of the CMF in dynamically formed BH LMXB population. 
 We do show, however, that in general the most likely companions are low-mass stars of $\rm M < 1 \rm \; M_{\odot}$ with the distribution 
 peaking at around $\rm \sim 0.9 \; M_{\odot}$ in all our models based on the Roche Lobe filling factor (see table~\ref{tab:first_table} and figure~\ref{fig:mass_distribution}). 
 Since we do not simulate the mass transfer itself, in reality 
 the donor masses in dynamically formed BH LMXBs should be even smaller. 
 This result is in agreement with the observed population of BH LMXBs for which donor masses span mostly in range 0.1--1.0 $\rm M_{\odot}$, 
 peaking at around $\rm \sim 0.6 \; M_{\odot}$ (see \citet{Wiktorowicz2014} and Fig.~\ref{fig:mass_distribution}).
 
\end{enumerate}

\section*{Acknowledgements}

We are grateful to the anonymous referee for careful reading of the manuscript and very helpful remarks.
We would like to thank Serena Repetto for useful comments and discussion.
We would also like to thank thousands of Universe@Home users who have provided
their personal computers for our simulations. Authors acknowledge support 
from the Polish NCN grant Sonata Bis 2 (DEC-2012/07/E/ST9/01360), the
Polish NCN grant OPUS (2015/19/B/ST9/01099) and 
the Polish NCN grant OPUS (2015/19/B/ST9/03188).




\bibliographystyle{mnras}
\bibliography{lmxbs_bib} 




%
%


\bsp	
\label{lastpage}
\end{document}